\documentclass[10pt]{article}
\usepackage{graphics,amssymb,epsfig,float,psfrag,axodraw2}
\usepackage{amsmath}
\usepackage{mathtools}
\usepackage{jheppub} % for details on the use of the package, please

\usepackage[T1]{fontenc} % if needed
\usepackage{bm}
\usepackage{natbib}
\usepackage{makecell}
\newcommand{\bea}{\begin{eqnarray}}
\newcommand{\eea}{\end{eqnarray}}
\newcommand{\beq}{\begin{eqnarray}}
\newcommand{\eeq}{\end{eqnarray}}

\newcommand{\bra}[1]{\langle#1|}

\def\spa#1.#2{\langle#1#2\rangle}
\def\spb#1.#2{[#1#2]}
\DeclarePairedDelimiter\za{\langle}{\rangle}
\DeclarePairedDelimiter\zb{[}{]}
\DeclareMathOperator*{\Res}{Res}
\def\tree{{\rm tree}}
\def\red{\color{Red}}
\def\blue{\color{Blue}}

\title{Shifts in BCFW method for QED}

\author[a]{Ke Li,}
\author[a]{Yuxin Liu,}
\author[a,b]{Qi-Shu Yan,}
\author[c]{Xiaoran Zhao}

\affiliation[a]{School of Physics Sciences, University of Chinese Academy of Sciences, Beijing 100039, China}
\affiliation[b]{Center for future high energy physics, Institute of High Energy Physics, Chinese Academy of Sciences, Beijing 100039, China}
\affiliation[c]{Dipartimento di Matematica e Fisica, Universit{\`a} di Roma Tre and \\
INFN, sezione di Roma Tre,\\
Via della Vasca Navale 84, Rome I-00146, Italy}

\emailAdd{yanqishu@ucas.ac.cn}
\emailAdd{xiaoran.zhao@uniroma3.it}

\abstract{ We study the application of BCFW recursion relations to the QED processes $0\to e^- e^+ n \gamma$. Based on 6-point amplitudes (both MHVA and NMHVA) computed from Feynman diagrams in the Berends-Giele gauge, we conduct a comprehensive study on all different shifts. Then we propose a new shift (LLYZ shift) which can lead to the full amplitudes for these processes and can have some realistic computation advantages. We compare the number of terms and the independent amplitudes of this novel shift with a few typical shifts.
}

\begin{document} 
\maketitle
\flushbottom

\section{Introduction}
Scattering amplitudes are the fundamental objects to theoretical predictions in perturbative quantum field theory.
Next generation colliders,
like the CEPC\cite{CEPC-SPPCStudyGroup:2015csa,An:2018dwb},
the CLIC\cite{Kalinowski:2018kdn,Charles:2018vfv,Linssen:2012hp,Aicheler:2012bya,Battaglia:2004mw,deBlas:2018mhx},
the FCC-ee\cite{Abada:2019zxq},
the ILC\cite{Fujii:2019zll},
and future muon colliders \cite{Delahaye:2019omf,Shiltsev:2019clx},
are expected to provide a clean environment for precision measurements,
as they adopt lepton beams.
Therefore, it is necessary to provide precise theoretical prediction,
where computing scattering amplitudes in QED are essential and necessary.

Traditionally, Feynman diagrams are adopted to compute the scattering amplitudes.
It originates from the Lagrangian formalism of quantum field theory,
where massless gauge bosons are represented as vector fields.
Such representation introduces unphysical degrees of freedom,
which must be removed by gauge fixing conditions.
As a result, individual Feynman diagrams are often gauge-dependent,
and physical results are obtained after summing all Feynman diagrams.
Such gauge redundancy often leads to rapidly growing number of Feynman diagrams as the number of external legs grows,
and thus the complexity increases significantly.

Powered by locality and unitarity, on-shell methods\cite{Britto:2004ap,Britto:2005fq} provide an alternative way which eliminate gauge dependency in the intermediate steps,
by constructing the full amplitudes via on-shell amplitudes only.
Initially it was discovered for pure Yang-Mills theory, then was extended to including quarks \cite{Schwinn:2007ee}, gravity \cite{Arkani-Hamed:2008bsc,Cheung:2008dn}\footnote{The reference \cite{Cheung:2008dn} has shown that if there is one gluon/graviton, there always exist good deformation.}, and SYM theory \cite{Elvang:2008na}.
Besides the analytical properties and validity of the on-shell recursion relations,
numerical studies have also be performed and compared with other methods.
In particular, in Ref. \cite{Dinsdale:2006sq} efficiency for color-ordered purely gluon amplitudes are studied,
and in Ref. \cite{Duhr:2006iq} the efficiency of evaluating full amplitude of gluon is studied. 

On the other hand, for QED, which is an Abelian theory,
such a method was firstly applied in Ref. \cite{Ozeren:2005mp}  to the process $e^+e^- \to n\gamma$. There the shift $[1, 5 \rangle$ (in our all-out conventions, it corresponds to $[5, 2\rangle$ shift) has been explored for the NMHVA for the process $e^- e^+ \to 4 \gamma$. The dressed version was proposed in Ref. \cite{Badger:2010eq} in order to obtain compact forms for amplitudes. Although with these progresses,
the core issue on which shift can be the simpler and more economic way for a realistic application is not addressed. In this work, we address this core issue and examine all possible shifts of the 
NMHVA for the process $e^- e^+ \to 4 \gamma$. Based on our observation, we propose a novel shift, which is a sum of a few shifts 
and can produce the correct amplitude, and we generalize our finding to the amplitudes of more general processes. 

In more details, we adopt the Feynman diagrams in Berends-Giele (BG) gauge \cite{Berends:1987me} to compute the helicity amplitudes of the process $ 0 \to e^+e^- 4\gamma$, $ 0 \to e^+e^- 5\gamma$,  and $ 0 \to e^+e^- 6\gamma$, . Then we use numerical approach to examine the equivalence of different shifts for the process $ 0 \to e^+e^- 4\gamma$ and then study the boundary terms for different shifts in the limit $z\to \infty$. Then we propose a novel shift (called LLYZ shift)  which could have some realistic computational advantages, though for each of the shifts the boundary term is non-vanishing. 
We also find that the shift $[1^-, \gamma^+ \rangle$ and the shift $[\gamma_i^-, \gamma_j^- \rangle$ of a pair of photons 
each with a negative helicity for the processes $0\to e^- e^+  \gamma_1^+ \cdots \gamma_n^+ \gamma_1^- \cdots \gamma_m^-$ with $1 < m \leq n$
can have manageable number of terms in the amplitudes, which should be considered in the realistic computations.

This paper is organized as follows. In section II, we briefly review the BCFW method. In section III, we present helicity amplitudes of the process $ 0 \to e^+e^- 4\gamma$, $ 0 \to e^+e^- 5\gamma$,  and $ 0 \to e^+e^- 6\gamma$. In section IV, we examine all shifts in the BCFW method for the NMHV amplitude of the process $ 0 \to e^+e^- 4\gamma$ and pay particular attentions to the boundary limits. Then we propose a so-called LLYZ shift. In section V, we prove that this shift can work for amplitudes of the general processes $0 \to e^- e^+ n \gamma$. In section VI, we examine the number of terms in the amplitudes and independent amplitudes for a few shifts.  Finally, we conclude this work with a few discussions.

\section{BCFW recursion relation}
Considering a tree-level amplitude $A=A(1,2,\cdots,i,\cdots,j,\cdots,n)$,
we can shift the momenta of particle $i$ and $j$ by the following shift on the spinors:
\begin{align}
 | \hat{i} \rangle = | i \rangle\,,\,\,| \hat{i} ] = | i ] + z | j ]\,,\quad
 | \hat{j} \rangle = | j \rangle - z | i\rangle\,,\,\, \ | \hat{j}] = | j ]\,.
 \label{shift}
\end{align}
We call this as $[i, j\rangle$ shift, which changes the spinors $[ i |$ and $| j \rangle$ while leave the spinors $[ j| $ and $| i \rangle$ unchanged.

With the above shift, it is clearly that momentum-conservation is preserved:
\begin{align}
    |\hat{j}\rangle[\hat{j}|+|\hat{i}\rangle[\hat{i}|=|j\rangle[j|+|i\rangle[i|
\end{align}
As a result, the amplitude is continued to the whole complex plane of $z$ as an analytic function: $A=A(1,2,\cdots,\hat{i},\cdots,\hat{j},\cdots,n)=A(z)$.
Considering the contour integral on the z plane,
\begin{align}
    I=\oint\frac{A(z)}{z}\,,
\end{align}
where the contour is large enough so that all finite singularities are inside the contour, as a result, we have
\begin{align}
0 =   \textrm{Res}_{z \to 0}\frac{A(z)}{z} +   \sum_{z\ne\infty, z\neq 0} \textrm{Res}_{z}\frac{A(z)}{z} + \textrm{Res}_{z\to \infty}\frac{A(z)}{z}\,.\label{eq:bcfw-residue}
\end{align}
For later usage, we can define $B= - \textrm{Res}_{z\to \infty}\frac{A(z)}{z}$ as the boundary term. $\textrm{Res}_{z \to 0}\frac{A(z)}{z} = A(0)$ is the original amplitude before the analytic continuation.
Then we can have
\bea
A(0) - B = - \sum_{z\ne\infty, z\neq 0} \textrm{Res}_{z}\frac{A(z)}{z} \,.
\eea

With an appropriate choice of a shift of two legs $ij$, we can make boundary term vanishing, i.e. $B=0$.  In such a case, 
it is known that the finite singularities of amplitude $A(z)$ only come from propagator denominators and then the amplitude can be factorized.
Thus, near the singular region $z \to z_I$, the amplitude can be factorized into left-hand and right-hand part which are connected by a propagator $\frac{1}{\hat{P}_L^2}$ and $\hat{P}_L=\sum_{k\in S}p_k$,
i.e. the amplitude can be put as
\begin{align}
    A(z) \xrightarrow{\hat{P}_L^2\to0} \sum_h \hat{A}_L(z_I, h) \,\, \frac{1}{\hat{P}_L^{2}} \,\,\hat{A}_R(z_I,-h) = - \frac{z_I}{z -z_I}  \sum_h \hat{A}_L(z_I, h) \,\, \frac{1}{P_L^{2}} \,\,\hat{A}_R(z_I,-h) \,,
\end{align}
where $L$ ($R$) denote the particles in the left-hand (right-hand) side, and $\hat{A}_L(z_I, h)$ ($\hat{A}_R(z_I, -h)$) are the sub-amplitude formed by particles of $L$ ($R$).

With the factorizibility near the pole regions and the analyticity given in \eqref{eq:bcfw-residue},
we can obtain the amplitude $A(0)$ in terms of on-shell amplitudes with less external legs:
\begin{align}
    A(0) =& -\sum_{i\in L, j\in R}\sum_{h}\Res_{\hat{P}_i^2(z)=0}\frac{1}{z} \hat{A}_L (z_I, h) \,\, \frac{1}{\hat{P}_L^2(z)} \,\,\hat{A}_R (z_I, -h)\\
    =& \sum_{i\in L,j\in R}\sum_{h} \hat{A}_L (z_I, h) \,\, \frac{1}{P_L^2} \,\,{\hat A}_R(z_I, -h) ,
\end{align}
which is the famous BCFW recursion relations. The advantage of assembling on-shell amplitudes into the full amplitudes lies in the fact that the subamplitudes are gauge independent and the calculation can be more efficient.

\section{\bf Helicity amplitudes for Process $ 0 \to e^- e^+ 4 \gamma $, $ 0 \to e^- e^+ 5 \gamma $ and $ 0 \to e^- e^+ 6\gamma $}
In this work, we will follow the conventions \cite{Elvang:2013cua} and use the all-out conventions for all amplitudes. It is useful to know that the mass dimension of the amplitudes. For the process $ 0 \to e^- e^+ \gamma \gamma \gamma \gamma$, the mass dimension of the amplitudes is equal to $4 - 6 = -2$.

  \begin{center}
 \vspace*{-.5cm}
\hspace*{-6.0cm}
\begin{picture}(300,110)(0,0)
\SetWidth{1.}
\SetScale{0.85}
\ArrowLine(120,90)(170,90)
\Text(170,90)[]{{\blue{\large $\bullet$}}}
\Text(170,60)[]{{\blue{\large $\bullet$}}}
\Text(170,30)[]{{\blue{\large $\bullet$}}}
\Text(170,0)[]{{\blue{\large $\bullet$}}}
\ArrowLine(170,90)(170,60)
\ArrowLine(170,60)(170,30)
\ArrowLine(170,30)(170,0)
\ArrowLine(170,0)(120,0)
%\ArrowLine(100,25)(130,50)
\Photon(170,90)(220,90){4}{5}
%\Line[arrow,arrowinset=0.3,arrowaspect=1,arrowwidth=1,arrowpos=1,
%arrowstroke=3](195,90)(200,90)
\Photon(170,60)(220,60){4}{5}
%\Line[arrow,arrowinset=0.3,arrowaspect=1,arrowwidth=1,arrowpos=1,
%arrowstroke=3](195,60)(200,60)
\Photon(170,30)(220,30){4}{5}
%\Line[arrow,arrowinset=0.3,arrowaspect=1,arrowwidth=1,arrowpos=1,
%arrowstroke=3](195,30)(200,30)
\Photon(170,0)(220,0){4}{5}
%\Line[arrow,arrowinset=0.3,arrowaspect=1,arrowwidth=1,arrowpos=1,
%arrowstroke=3](195,0)(200,0)
\Text(155,85)[]{$2^+$}
\Text(155,5)[]{$1^-$}
\Text(210,80)[]{$3^+$}
\Text(210,50)[]{$4^+$}
\Text(210,20)[]{$5^+$}
\Text(210,-10)[]{$6^-$}
\Text(170,-15)[]{{\red{\bf 1.a)}}}
\hspace*{1cm}
\SetWidth{1.}
\ArrowLine(270,90)(320,90)
\Text(320,90)[]{{\blue{\large $\bullet$}}}
\Text(320,60)[]{{\blue{\large $\bullet$}}}
\Text(320,30)[]{{\blue{\large $\bullet$}}}
\Text(320,0)[]{{\blue{\large $\bullet$}}}
\ArrowLine(320,90)(320,60)
\ArrowLine(320,60)(320,30)
\ArrowLine(320,30)(320,0)
\ArrowLine(320,0)(270,0)
%\ArrowLine(100,25)(130,50)
\Photon(320,90)(370,90){4}{5}
%\Line[arrow,arrowinset=0.3,arrowaspect=1,arrowwidth=1,arrowpos=1,
%arrowstroke=3](345,90)(350,90)
\Photon(320,60)(370,60){4}{5}
%\Line[arrow,arrowinset=0.3,arrowaspect=1,arrowwidth=1,arrowpos=1,
%arrowstroke=3](345,60)(350,60)
\Photon(320,30)(370,30){4}{5}
%\Line[arrow,arrowinset=0.3,arrowaspect=1,arrowwidth=1,arrowpos=1,
%arrowstroke=3](345,30)(350,30)
\Photon(320,0)(370,0){4}{5}
%\Line[arrow,arrowinset=0.3,arrowaspect=1,arrowwidth=1,arrowpos=1,
%arrowstroke=3](345,0)(350,0)
\Text(305,85)[]{$2^+$}
\Text(305,5)[]{$1^-$}
\Text(360,80)[]{$3^+$}
\Text(360,50)[]{$4^+$}
\Text(360,20)[]{$5^-$}
\Text(360,-10)[]{$6^-$}
\Text(320,-15)[]{{\red{\bf 1.b)}}}
\end{picture}
\end{center}
\vspace*{0.5cm}
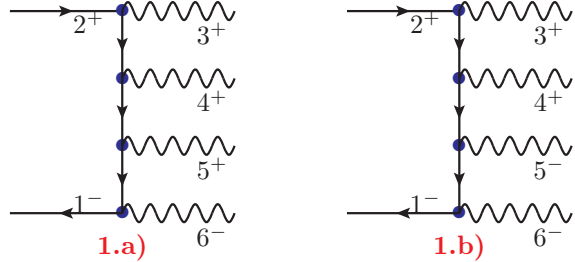
\centerline{\it Figure 1: Two non-vanishing Feynman Diagrams of $0 \to e^-(1) e^+(2) \gamma(3) \gamma(4) \gamma(5) \gamma(6) $ are shown. \label{figure1}}

 The helicity amplitudes of $ 0 \to e^- e^+ \gamma \gamma \gamma \gamma$ include 24 Feynman diagrams, there are only two types of independent helicity amplitudes: one is the MHV and the other is NMHV. In Figure 1, two Feynman diagrams for each of these two types of amplitudes are presented.

The MHV amplitude can be computed in the Feynman diagram method. In the BG gauge \cite{Berends:1987me} , the spinors of a photon can be expressed as
\bea
\epsilon^{-}(i) = \frac{|i\rangle [q_i|}{\spb{i}.{q_i}}\,\,, \epsilon^{+}(i) = \frac{|q_i \rangle [i|}{\spa{i}.{q_i}}\,,
\eea
where momentum $q_i$ denotes reference momentum. By choosing the reference momenta as $q_6 =2$ and $q_3=q_4=q_5=1$ and computing 6 nonvanishing Feynman diagrams as shown in Figure 1.a) (where diagrams with all permutations of (3,4,5) should be summed), we find that the MHV amplitudes can be put as given below
\bea
A^\tree_6(1_e^-,2_{\bar{e}}^+,3_{\gamma}^+,4_{\gamma}^+,5_{\gamma}^+, 6_{\gamma}^-) = Q_e^4\frac{{\spa1.2}^2 {\spa1.6}^2}{\spa1.3 \spa2.3 \spa1.4  \spa2.4 \spa1.5  \spa2.5}\,. \label{aee4a}
\eea
Obviously, the amplitude is invariant under the exchange $ 3 \leftrightarrow 4$, or $4 \leftrightarrow 5$, or $5 \leftrightarrow 3$, which is expected due to the bosonic nature of photons. 
Such an amplitude can also be obtained by using the BCFW method and taking the $[ 1, 2 \rangle $ shift. Then by utilizing the boson exchanging symmetry we arrive at the amplitude 
 \bea
 \frac{\spa1.6^2 }{\spa2.3 \spa2.4 \spa2.5} \left \{  \frac{\spa2.3^2}{\spa1.3 \spa4.3 \spa5.3} + \frac{\spa2.4^2}{\spa1.4 \spa3.4 \spa5.4} + \frac{\spa2.5^2}{\spa1.5 \spa3.5 \spa4.5}  \right \}\,.\label{bcfwmhv}
 \eea
After combining all terms in the bracket, it is found that 
 \bea
 \frac{\spa2.3^2}{\spa1.3 \spa4.3 \spa5.3} +  \frac{\spa2.4^2}{\spa1.4 \spa3.4 \spa5.4} + \frac{\spa2.5^2}{\spa1.5 \spa3.5 \spa4.5}  = \frac{\spa1.2^2}{\spa1.3 \spa1.4 \spa1.5}\,.
 \eea
Then we arrive at the result given in Eq. (\ref{aee4a}) from Eq. (\ref{bcfwmhv}). The MHV helicity amplitudes with more photons of the process $ 0 \to e^-(1^-) e^+(2^+) \gamma^-(3^-) \gamma^+(4^+)  \cdots \gamma^+(n^+) $ can be conjectured as
\bea
A^\tree_{n}(1_e^-,2_{\bar{e}}^+,3_{\gamma}^-,4_{\gamma}^+,5_{\gamma}^+, 6_{\gamma}^+, \cdots, n_{\gamma}^+) \rightarrow A^\tree_4(1_e^-,2_{\bar{e}}^+,3_{\gamma}^-,4_{\gamma}^+) \frac{\spa1.2}{\spa1.5 \spa2.5} \cdots  \frac{\spa1.2}{\spa1.n \spa2.n}\,,
\label{nrc}
\eea
which have the correct mass dimensions, the correct helicity index for each spinor, and the boson exchanging symmetries among indices ($4, 5, \cdots,$ and $n$).

By using the BCFW method and taking the shift $[ 1, 2 \rangle $, we can obtain the amplitude in the following form 
\bea
\frac{\spa1.3^2 }{\spa2.4 \cdots \spa2.n } \left\{ \frac{\spa2.4^{n-4} }{\spa1.4 \spa5.4 \spa6.4 \cdots \spa{n}.4} + \cdots + \frac{\spa2.{n}^{n-4}} {\spa1.n  \spa4.n \spa5.n \cdots \spa{(n-1)}.n} \right \}\,, \label{bcfwnp}
\eea
which keeps the boson exchanging symmetries manifest. By using induction, we can arrive at the result given in Eq. (\ref{nrc}) from Eq. (\ref{bcfwnp}). Therefore, the MHV amplitudes of QED can be elegantly put as
\bea
A^\tree_{n}(1_e^-,2_{\bar{e}}^+,3_{\gamma}^-,4_{\gamma}^+,5_{\gamma}^+, 6_{\gamma}^+, \cdots, n_{\gamma}^+) = Q_e^{n-2} \frac{{\spa1.3}^2 {\spa1.2}^{n-4}}{\spa1.4 \spa2.4 \spa1.5 \spa2.5  \cdots \spa1.n \spa2.n }\,,
\eea
which has only one term and has been obtained in \cite{Berends:1987me}.

The NMHV amplitude can be computed in the Feynman diagram method in the BG gauge.  By choosing the gauge with $q_3 = q_4 =1$ and $q_5 = q_6 = 2$ and computing 8 non-vanishing Feynman diagrams, we arrive at the amplitude of NMHV which can be put as
\bea
A^\tree_6(1_e^-,2_{\bar{e}}^+,3_{\gamma}^+,4_{\gamma}^+,5_{\gamma}^-, 6_{\gamma}^-) =Q_e^4 \left \{ \frac{1 }{\spa1.3 \spa2.3 \spa1.4 \spa2.4 \spb1.5 \spb2.5 \spb1.6 \spb2.6 } S^1_6 + S^2_6 \right \} \label{nmhv}\,,
\eea
where the term $S_6^1$ computed from the four Feynman diagrams given in Figure 2, 
which are given as 
\bea
S_6^1 =  \spa1.2 \spb1.2 \,\,  \langle 1 | k \!\!\!\slash_{156} | 2] \,\, S_{156} = \spa1.2 \spb1.2 \,\,  \langle 1 | 5+6 | 2] \,\, S_{156} \,,
\label{eeaaaa21}
\eea

  \begin{center}
 \vspace*{-1.5cm}
\hspace*{-4.0cm}
\begin{picture}(300,120)(0,0)
\SetWidth{1.}
\SetScale{0.8}
\ArrowLine(20,90)(70,90)
\Text(70,90)[]{{\blue{\large $\bullet$}}}
\Text(70,60)[]{{\blue{\large $\bullet$}}}
\Text(70,30)[]{{\blue{\large $\bullet$}}}
\Text(70,0)[]{{\blue{\large $\bullet$}}}
\ArrowLine(70,90)(70,60)
\ArrowLine(70,60)(70,30)
\ArrowLine(70,30)(70,0)
\ArrowLine(70,0)(20,0)
%\ArrowLine(100,25)(130,50)
\Photon(70,90)(120,90){4}{5}
%\Line[arrow,arrowinset=0.3,arrowaspect=1,arrowwidth=1,arrowpos=1,
%arrowstroke=3](95,90)(100,90)
\Photon(70,60)(120,60){4}{5}
%\Line[arrow,arrowinset=0.3,arrowaspect=1,arrowwidth=1,arrowpos=1,
%arrowstroke=3](95,60)(100,60)
\Photon(70,30)(120,30){4}{5}
%\Line[arrow,arrowinset=0.3,arrowaspect=1,arrowwidth=1,arrowpos=1,
%arrowstroke=3](95,30)(100,30)
\Photon(70,0)(120,0){4}{5}
%\Line[arrow,arrowinset=0.3,arrowaspect=1,arrowwidth=1,arrowpos=1,
%arrowstroke=3](95,0)(100,0)
\Text(55,85)[]{$2^+$}
\Text(55,5)[]{$1^-$}
\Text(110,80)[]{$3^+$}
\Text(110,50)[]{$4^+$}
\Text(110,20)[]{$5^-$}
\Text(110,-10)[]{$6^-$}
\Text(70,-15)[]{{\red{\bf 2.a)}}}
%\hspace*{1cm}
\ArrowLine(140,90)(190,90)
\Text(190,90)[]{{\blue{\large $\bullet$}}}
\Text(190,60)[]{{\blue{\large $\bullet$}}}
\Text(190,30)[]{{\blue{\large $\bullet$}}}
\Text(190,0)[]{{\blue{\large $\bullet$}}}
\ArrowLine(190,90)(190,60)
\ArrowLine(190,60)(190,30)
\ArrowLine(190,30)(190,0)
\ArrowLine(190,0)(140,0)
%\ArrowLine(100,25)(130,50)
\Photon(190,90)(240,90){4}{5}
%\Line[arrow,arrowinset=0.3,arrowaspect=1,arrowwidth=1,arrowpos=1,
%arrowstroke=3](215,90)(220,90)
\Photon(190,60)(240,60){4}{5}
%\Line[arrow,arrowinset=0.3,arrowaspect=1,arrowwidth=1,arrowpos=1,
%arrowstroke=3](215,60)(220,60)
\Photon(190,30)(240,30){4}{5}
%\Line[arrow,arrowinset=0.3,arrowaspect=1,arrowwidth=1,arrowpos=1,
%arrowstroke=3](215,30)(220,30)
\Photon(190,0)(240,0){4}{5}
%\Line[arrow,arrowinset=0.3,arrowaspect=1,arrowwidth=1,arrowpos=1,
%arrowstroke=3](215,0)(220,0)
\Text(175,85)[]{$2^+$}
\Text(175,5)[]{$1^-$}
\Text(230,80)[]{$3^+$}
\Text(230,50)[]{$4^+$}
\Text(230,20)[]{$6^-$}
\Text(230,-10)[]{$5^-$}
\Text(190,-15)[]{{\red{\bf 2.b)}}}
\ArrowLine(260,90)(310,90)
\Text(310,90)[]{{\blue{\large $\bullet$}}}
\Text(310,60)[]{{\blue{\large $\bullet$}}}
\Text(310,30)[]{{\blue{\large $\bullet$}}}
\Text(310,0)[]{{\blue{\large $\bullet$}}}
\ArrowLine(310,90)(310,60)
\ArrowLine(310,60)(310,30)
\ArrowLine(310,30)(310,0)
\ArrowLine(310,0)(260,0)
%\ArrowLine(100,25)(130,50)
\Photon(310,90)(360,90){4}{5}
%\Line[arrow,arrowinset=0.3,arrowaspect=1,arrowwidth=1,arrowpos=1,
%arrowstroke=3](335,90)(340,90)
\Photon(310,60)(360,60){4}{5}
%\Line[arrow,arrowinset=0.3,arrowaspect=1,arrowwidth=1,arrowpos=1,
%arrowstroke=3](335,60)(340,60)
\Photon(310,30)(360,30){4}{5}
%\Line[arrow,arrowinset=0.3,arrowaspect=1,arrowwidth=1,arrowpos=1,
%arrowstroke=3](335,30)(340,30)
\Photon(310,0)(360,0){4}{5}
%\Line[arrow,arrowinset=0.3,arrowaspect=1,arrowwidth=1,arrowpos=1,
%arrowstroke=3](335,0)(340,0)
\Text(295,85)[]{$2^+$}
\Text(295,5)[]{$1^-$}
\Text(350,80)[]{$4^+$}
\Text(350,50)[]{$3^+$}
\Text(350,20)[]{$5^-$}
\Text(350,-10)[]{$6^-$}
\Text(310,-15)[]{{\red{\bf 2.c)}}}
\ArrowLine(380,90)(430,90)
\Text(430,90)[]{{\blue{\large $\bullet$}}}
\Text(430,60)[]{{\blue{\large $\bullet$}}}
\Text(430,30)[]{{\blue{\large $\bullet$}}}
\Text(430,0)[]{{\blue{\large $\bullet$}}}
\ArrowLine(430,90)(430,60)
\ArrowLine(430,60)(430,30)
\ArrowLine(430,30)(430,0)
\ArrowLine(430,0)(380,0)
%\ArrowLine(100,25)(130,50)
\Photon(430,90)(480,90){4}{5}
%\Line[arrow,arrowinset=0.3,arrowaspect=1,arrowwidth=1,arrowpos=1,
%arrowstroke=3](455,90)(460,90)
\Photon(430,60)(480,60){4}{5}
%\Line[arrow,arrowinset=0.3,arrowaspect=1,arrowwidth=1,arrowpos=1,
%arrowstroke=3](455,60)(460,60)
\Photon(430,30)(480,30){4}{5}
%\Line[arrow,arrowinset=0.3,arrowaspect=1,arrowwidth=1,arrowpos=1,
%arrowstroke=3](455,30)(460,30)
\Photon(430,0)(480,0){4}{5}
%\Line[arrow,arrowinset=0.3,arrowaspect=1,arrowwidth=1,arrowpos=1,
%arrowstroke=3](455,0)(460,0)
\Text(415,85)[]{$2^+$}
\Text(415,5)[]{$1^-$}
\Text(470,80)[]{$4^+$}
\Text(470,50)[]{$3^+$}
\Text(470,20)[]{$6^-$}
\Text(470,-10)[]{$5^-$}
\Text(430,-15)[]{{\red{\bf 2.d)}}}
\end{picture}
\end{center}
\vspace*{0.5cm}
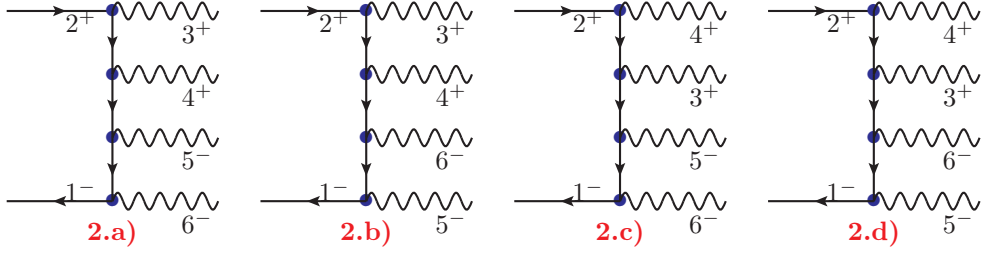
\centerline{\it Figure 2: Four Feynman Diagrams contributing to the term $S^1_6$ are shown. \label{figure2}} 

When we consider the soft limits below, it is noted that the following two relations are important and crucial 
\bea
\langle 1 | {k \!\!\!\slash} _{156} | 2] = - \langle 1 | {k \!\!\!\slash}_{234} | 2]
\eea
and 
\bea
S_{234} = S_{156} = (k_2 + k_3 + k_4)^2 = (k_1 + k_5 + k_6)^2\,,
\eea
which are the direct results of momentum conservation $k_1 +k_2 +k_3 +k_4 +k_5+k_6=0 $. These two relations are also helpful to understand the symmetric features of $S^1_6$. Obviously, $S^1_6$ is invariant under the exchange transformations $3 \leftrightarrow 4$ and $5 \leftrightarrow 6$. 

%The term proportional to $S_6^1$ can also be obtained by combining two off-shell amplitudes given in Eq. (\ref{oa3l}) and Eq. (\ref{oa3r}) in Appendix \ref{appb}, respectively. It should be pointed out that a two-point function in the form of spinor ${k\!\!\!\slash}_{156}$ should be added between these two off-shell amplitudes. Then by using the energy-momentum conservation, ${k\!\!\!\slash}_{234} = - {k\!\!\!\slash}_{156}$ and ${k\!\!\!\slash}_{156} {k\!\!\!\slash}_{156} = S_{156}$, it is straightforward to arrive at the term proportional to $S^1_6$. It is noticed that similar terms for the amplitudes with more points can be derived.

  \begin{center}
 \vspace*{-1.5cm}
\hspace*{-4.0cm}
\begin{picture}(300,120)(0,0)
\SetWidth{1.}
\SetScale{0.8}
\ArrowLine(20,90)(70,90)
\Text(70,90)[]{{\blue{\large $\bullet$}}}
\Text(70,60)[]{{\blue{\large $\bullet$}}}
\Text(70,30)[]{{\blue{\large $\bullet$}}}
\Text(70,0)[]{{\blue{\large $\bullet$}}}
\ArrowLine(70,90)(70,60)
\ArrowLine(70,60)(70,30)
\ArrowLine(70,30)(70,0)
\ArrowLine(70,0)(20,0)
%\ArrowLine(100,25)(130,50)
\Photon(70,90)(120,90){4}{5}
%\Line[arrow,arrowinset=0.3,arrowaspect=1,arrowwidth=1,arrowpos=1,
%arrowstroke=3](95,90)(100,90)
\Photon(70,60)(120,60){4}{5}
%\Line[arrow,arrowinset=0.3,arrowaspect=1,arrowwidth=1,arrowpos=1,
%arrowstroke=3](95,60)(100,60)
\Photon(70,30)(120,30){4}{5}
%\Line[arrow,arrowinset=0.3,arrowaspect=1,arrowwidth=1,arrowpos=1,
%arrowstroke=3](95,30)(100,30)
\Photon(70,0)(120,0){4}{5}
%\Line[arrow,arrowinset=0.3,arrowaspect=1,arrowwidth=1,arrowpos=1,
%arrowstroke=3](95,0)(100,0)
\Text(55,85)[]{$2^+$}
\Text(55,5)[]{$1^-$}
\Text(110,80)[]{$3^+$}
\Text(110,50)[]{$6^-$}
\Text(110,20)[]{$4^+$}
\Text(110,-10)[]{$5^-$}
\Text(70,-15)[]{{\red{\bf 3.a)}}}
%\hspace*{1cm}
\ArrowLine(140,90)(190,90)
\Text(190,90)[]{{\blue{\large $\bullet$}}}
\Text(190,60)[]{{\blue{\large $\bullet$}}}
\Text(190,30)[]{{\blue{\large $\bullet$}}}
\Text(190,0)[]{{\blue{\large $\bullet$}}}
\ArrowLine(190,90)(190,60)
\ArrowLine(190,60)(190,30)
\ArrowLine(190,30)(190,0)
\ArrowLine(190,0)(140,0)
%\ArrowLine(100,25)(130,50)
\Photon(190,90)(240,90){4}{5}
%\Line[arrow,arrowinset=0.3,arrowaspect=1,arrowwidth=1,arrowpos=1,
%arrowstroke=3](215,90)(220,90)
\Photon(190,60)(240,60){4}{5}
%\Line[arrow,arrowinset=0.3,arrowaspect=1,arrowwidth=1,arrowpos=1,
%arrowstroke=3](215,60)(220,60)
\Photon(190,30)(240,30){4}{5}
%\Line[arrow,arrowinset=0.3,arrowaspect=1,arrowwidth=1,arrowpos=1,
%arrowstroke=3](215,30)(220,30)
\Photon(190,0)(240,0){4}{5}
%\Line[arrow,arrowinset=0.3,arrowaspect=1,arrowwidth=1,arrowpos=1,
%arrowstroke=3](215,0)(220,0)
\Text(175,85)[]{$2^+$}
\Text(175,5)[]{$1^-$}
\Text(230,80)[]{$3^+$}
\Text(230,50)[]{$5^-$}
\Text(230,20)[]{$4^+$}
\Text(230,-10)[]{$6^-$}
\Text(190,-15)[]{{\red{\bf 3.b)}}}
\ArrowLine(260,90)(310,90)
\Text(310,90)[]{{\blue{\large $\bullet$}}}
\Text(310,60)[]{{\blue{\large $\bullet$}}}
\Text(310,30)[]{{\blue{\large $\bullet$}}}
\Text(310,0)[]{{\blue{\large $\bullet$}}}
\ArrowLine(310,90)(310,60)
\ArrowLine(310,60)(310,30)
\ArrowLine(310,30)(310,0)
\ArrowLine(310,0)(260,0)
%\ArrowLine(100,25)(130,50)
\Photon(310,90)(360,90){4}{5}
%\Line[arrow,arrowinset=0.3,arrowaspect=1,arrowwidth=1,arrowpos=1,
%arrowstroke=3](335,90)(340,90)
\Photon(310,60)(360,60){4}{5}
%\Line[arrow,arrowinset=0.3,arrowaspect=1,arrowwidth=1,arrowpos=1,
%arrowstroke=3](335,60)(340,60)
\Photon(310,30)(360,30){4}{5}
%\Line[arrow,arrowinset=0.3,arrowaspect=1,arrowwidth=1,arrowpos=1,
%arrowstroke=3](335,30)(340,30)
\Photon(310,0)(360,0){4}{5}
%\Line[arrow,arrowinset=0.3,arrowaspect=1,arrowwidth=1,arrowpos=1,
%arrowstroke=3](335,0)(340,0)
\Text(295,85)[]{$2^+$}
\Text(295,5)[]{$1^-$}
\Text(350,80)[]{$4^+$}
\Text(350,50)[]{$6^-$}
\Text(350,20)[]{$3^+$}
\Text(350,-10)[]{$5^-$}
\Text(310,-15)[]{{\red{\bf 3.c)}}}
\ArrowLine(380,90)(430,90)
\Text(430,90)[]{{\blue{\large $\bullet$}}}
\Text(430,60)[]{{\blue{\large $\bullet$}}}
\Text(430,30)[]{{\blue{\large $\bullet$}}}
\Text(430,0)[]{{\blue{\large $\bullet$}}}
\ArrowLine(430,90)(430,60)
\ArrowLine(430,60)(430,30)
\ArrowLine(430,30)(430,0)
\ArrowLine(430,0)(380,0)
%\ArrowLine(100,25)(130,50)
\Photon(430,90)(480,90){4}{5}
%\Line[arrow,arrowinset=0.3,arrowaspect=1,arrowwidth=1,arrowpos=1,
%arrowstroke=3](455,90)(460,90)
\Photon(430,60)(480,60){4}{5}
%\Line[arrow,arrowinset=0.3,arrowaspect=1,arrowwidth=1,arrowpos=1,
%arrowstroke=3](455,60)(460,60)
\Photon(430,30)(480,30){4}{5}
%\Line[arrow,arrowinset=0.3,arrowaspect=1,arrowwidth=1,arrowpos=1,
%arrowstroke=3](455,30)(460,30)
\Photon(430,0)(480,0){4}{5}
%\Line[arrow,arrowinset=0.3,arrowaspect=1,arrowwidth=1,arrowpos=1,
%arrowstroke=3](455,0)(460,0)
\Text(415,85)[]{$2^+$}
\Text(415,5)[]{$1^-$}
\Text(470,80)[]{$4^+$}
\Text(470,50)[]{$5^-$}
\Text(470,20)[]{$3^+$}
\Text(470,-10)[]{$6^-$}
\Text(430,-15)[]{{\red{\bf 3.d)}}}
\end{picture}
\end{center}
\vspace*{0.5cm}
\centerline{\it Figure 3: Four non-vanishing Feynman Diagrams contributing to the term $S^2_6$ are shown.\label{fig:fd3}} 

The term $S_6^2$ can be put in a symmetric form from the four Feynman diagrams given in Figure 3 as
\bea
S_6^2 & = &\frac{1}{ \spa1.4  \spb2.6} \frac{ \spa1.5 \spb2.3}{\spb1.5 \spa2.3 } \frac{\langle 6 |{k \!\!\!\slash}_{236} | 4 ] }{S_{236}} + \frac{1}{ \spa1.4  \spb2.5} \frac{ \spa1.6 \spb2.3}{\spb1.6 \spa2.3 } \frac{ \langle 5 |{k \!\!\!\slash}_{235} | 4 ]  }{S_{235}} \nonumber \\
 & +&  \frac{1}{ \spa1.3  \spb2.6} \frac{ \spa1.5 \spb2.4}{\spb1.5 \spa2.4 } \frac{ \langle 6 |{k \!\!\!\slash}_{246} | 3 ] }{S_{246}} + \frac{1}{ \spa1.3  \spb2.5} \frac{ \spa1.6 \spb2.4}{\spb1.6 \spa2.4 } \frac{ \langle 5 |{k \!\!\!\slash}_{245} | 3 ] }{S_{245}}\,.
 \eea
Using the fact $\langle i | i+j+k | j]=\langle i |k | j]$, we can present the term $S_6^2$ as
\bea
S_6^2 & = &\frac{1}{ \spa1.4  \spb2.6} \frac{ \spa1.5 \spb2.3}{\spb1.5 \spa2.3 } \frac{\langle 6 |2+3 | 4 ] }{S_{236}} + \frac{1}{ \spa1.4  \spb2.5} \frac{ \spa1.6 \spb2.3}{\spb1.6 \spa2.3 } \frac{ \langle 5 |2+3| 4 ]  }{S_{235}} \nonumber \\
 & +&  \frac{1}{ \spa1.3  \spb2.6} \frac{ \spa1.5 \spb2.4}{\spb1.5 \spa2.4 } \frac{ \langle 6 |2+4 | 3 ] }{S_{246}} + \frac{1}{ \spa1.3  \spb2.5} \frac{ \spa1.6 \spb2.4}{\spb1.6 \spa2.4 } \frac{ \langle 5 |2+4 | 3 ] }{S_{245}}\,\\
 &=& [P(5,6)] [P(3,4)]  \frac{1}{ \spa1.4  \spb2.6} \frac{ \spa1.5 \spb2.3}{\spb1.5 \spa2.3 } \frac{\langle 6 |2+3 | 4 ] }{S_{236}} \,,
\eea
where we have use the permutation group symbols $[P(5,6)]$ and $[P(3,4)]$ to simplify the result. $[P(i,j)]$ denotes the sum of all group elements of permutation group of two objects, i.e. $[P(i,j)] = E + P(i,j)$. Since photons $5$ and $6$ ($3$ and $4$) have the same helicity, there exists such boson exchanging symmetry.

In total, the amplitude given in Eq. (\ref{nmhv}) is manifestly invariant under the exchange $3 \leftrightarrow 4$ and $5 \leftrightarrow 6$, as required by the boson exchanging symmetry. It is noticed that $S^2_6$ vanishes when any one of photons goes to its soft limit. With this property, the form given in Eq. (\ref{nmhv}) has very nice features when we consider all those soft and collinear limits.

It should be mentioned that although in reference \cite{Kleiss:1986qc} a very general form for the amplitudes with arbitrary number of photons has been formulated in terms of Feynman diagrams, the number of terms increases as $n!$ ($n$ is the number of photons). In contrast, in the BG gauge, there is only one compact term for the MHV amplitudes as shown above. Although there is no general form of NMHV and N$^2$MHV (or higher) amplitudes for more photons in BG gauge, the actual forms of the NMHV or higher order amplitudes are dependent upon the helicity sequences (or spin chain of photons). Once the helicity sequences are specified, it it straightforward to write down the amplitudes in terms of spinor brackets and permutation groups. Below we present the amplitudes of $0 \to e^- e^+ 5 \gamma$ and $0 \to e^- e^+ 6\gamma$ for later reference. 

There is a fact of which should be reminded. In the total amplitude, in order to realize the boson exchanging symmetries, all possible permutations of photons with the same helicities have to be taken into account. For these helicity configurations, it is noticed that both end particles denoted as $\bar{f}(2^+)$ and $f(1^-)$ are fixed, the helicities of the photons adjoined to fermion particles are also fixed in the BG gauge. These facts constrain all allowed helicity configurations. For the NMHV 6-point amplitude, there are only two helicity configurations, i.e. $+++---$,  and $++-+--$. For the NMHV 7-point amplitude, there are only three helicity configurations allowed, i.e.  $++++---$, $+++-+--$ and $++-++--$. And for the NMHV 8-point amplitude, there are only four helicity configurations allowed, i.e.  $+++++---$, $++++-+--$, $+++-++--$, and $++-+++--$. For N$^2$MHV of 8-point amplitude, there are only six helicity configurations allowed, i.e. $++++----$, $+++-+---$, $++-++---$, $+++--+--$, $++-+--+--$, and $++--++--$. For more general amplitudes of the processes  $0\to e^- e^+ \gamma_1^+ \cdots \gamma_n^+ \gamma_1^- \cdots \gamma_m^-$ with $ m \leq n$, the number of allowed helicity configurations is given as
\bea
N_S = \frac{(m+n-2)!}{(m-1)! \,\,(n-1)! }\,,
\eea
which is much less than the number of amplitudes constructed directly from Feynman diagram ( which should be counted as $(m+n)!$). 

It is noteworthy that terms calculated in the helicity amplitude can be combined into a form with less number of terms, like there is only one term for MHV amplitudes. For helicity amplitudes beyond the MHVA, less number of terms can be achieved just like the helicity amplitudes given in Eq. (3.17), Eq. (3.22) and Eq. (3.27). While in contrast, if we construct the amplitude directly from Feynman diagrams, naively there are n! terms no matter whatever a helicity structure is assumed.

Then, the NMHV amplitude of $A^\tree_7(1_e^-,2_{\bar{e}}^+,3_{\gamma}^+,4_{\gamma}^+,5_{\gamma}^+, 6_{\gamma}^-, 7_{\gamma}^-)$ can be computed efficiently by using the off-shell amplitude method in the BG gauge. Using the Feynman diagrams given in Figure 4 as examples and using the boson exchanging symmetries for photons which have the same helicities, we can obtain the total amplitude. In terms of helicity configrations, the total amplitude can be organized into the following form
 \bea
A^\tree_7(1_e^-,2_{\bar{e}}^+,3_{\gamma}^+,4_{\gamma}^+,5_{\gamma}^+, 6_{\gamma}^-, 7_{\gamma}^-) &=& Q_e^5 \left \{  S^1_7 + S^2_7 + S_7^3 \right \} \label{nmhv7t}\,, 
\eea
where each $S^i_7$ term is given below
\bea
S^1_7 &= & \frac{\spa1.2^2 \spb1.2 \,\, \langle 1 | 6 + 7 | 2] \,\, S_{167} }{\prod_{i=3,4,5} \spa1.i \spa2.i  \prod_{j=6,7} \spb1.j \spb2.j  }\,,\\
S^2_7 & = &\frac{\spa1.2}{\prod_{i=3,4,5}\spa1.i } [P(6,7)] \left [  \frac{\spa1.7}{\spb1.7 \spb2.6}  [P(34,5)] \frac{ \langle 1 | 3+4| 2 ] }{ \spa2.3 \spa2.4 }   \frac{[5 | 1 + 7 | 6 \rangle}{S_{157}}  \right ]  \label{a72} \,,\\
S^3_7&=& [P(6,7)] \frac{\spa1.7^2}{\spb1.7 \spb2.6 } \left [ [P(3,45)]   \frac{\spb2.3}{\spa2.3} \frac{1}{\spa1.4 \spa1.5 S_{236}} [P(4,5)] \frac{\spb5.7 [4 | 2 + 3 | 6 \rangle}{S_{157}} \right ] \label{a73}\,,
\eea
where $S_7^1$ includes the contribution of diagrams with the helicity configuration $++++---$ as given in Figure 4.a, $S^2_7$ includes the contribution of diagrams with the helicity configuration $+++-+--$ as given in Figure 4.b, and $S^3_7$ includes the contribution of diagrams with the helicity configuration $++-++--$ as shown in Figure 4.c. The symbol $[P(34,5)]$ denotes the sum of the subset of permutation group $E + P_{35} + P_{45}$, since the symmetry between $3$ and $4$ is realized and only the symmetry with $5$ needs to be restored, as shown in Figure 4.b. The symbol $[P(34,5)]$ acts the terms in its right hand side, i.e. after the action, the terms will be changed as the sum of three terms. For example,
\bea
[P(34,5)] \frac{ \langle 1 | 3+4| 2 ] }{ \spa2.3 \spa2.4 }   \frac{[5 | 1 + 7 | 6 \rangle}{S_{157}} &=& \frac{ \langle 1 | 3+4| 2 ] }{ \spa2.3 \spa2.4 }   \frac{[5 | 1 + 7 | 6 \rangle}{S_{157}}  \nonumber \\
& + & \frac{ \langle 1 | 4+5| 2 ] }{ \spa2.4 \spa2.5 }   \frac{[3 | 1 + 7 | 6 \rangle}{S_{137}} +  \frac{ \langle 1 | 3+5| 2 ] }{ \spa2.3 \spa2.5 }   \frac{[4 | 1 + 7 | 6 \rangle}{S_{147}} \,.
\eea
Similarly, the symbol $[P(3,45)]$ denotes the sum of $E + P_{34} + P_{35}$. When $[P(3,45)] [P(4,5)]$, we arrive at $[P(3,45)] [P(4,5)] = [P(3,4,5)]$ which denotes the sum of all allowed permutations of three objects, i.e. $[P(3,4,5)] = E + P_{34} + P_{35}+ P_{45} + P_{345} + P_{345}^2$.

 \begin{center}
 \vspace*{-0.5cm}
\hspace*{1.0cm}
\begin{picture}(360,120)(0,0)
\SetWidth{1.}
\SetScale{0.7}
\ArrowLine(0,120)(50,120)
\Text(50,120)[]{{\blue{\large $\bullet$}}}
\Text(50,90)[]{{\blue{\large $\bullet$}}}
\Text(50,60)[]{{\blue{\large $\bullet$}}}
\Text(50,30)[]{{\blue{\large $\bullet$}}}
\Text(50,0)[]{{\blue{\large $\bullet$}}}
\ArrowLine(50,120)(50,90)
\ArrowLine(50,90)(50,60)
\ArrowLine(50,60)(50,30)
\ArrowLine(50,30)(50,0)
\ArrowLine(50,0)(0,0)
%\ArrowLine(100,25)(130,50)
\Photon(50,120)(100,120){4}{5}
%\Line[arrow,arrowinset=0.3,arrowaspect=1,arrowwidth=1,arrowpos=1,
%arrowstroke=3](95,90)(100,90)
\Photon(50,90)(100,90){4}{5}
%\Line[arrow,arrowinset=0.3,arrowaspect=1,arrowwidth=1,arrowpos=1,
%arrowstroke=3](95,60)(100,60)
\Photon(50,60)(100,60){4}{5}
%\Line[arrow,arrowinset=0.3,arrowaspect=1,arrowwidth=1,arrowpos=1,
%arrowstroke=3](95,30)(100,30)
\Photon(50,30)(100,30){4}{5}
\Photon(50,0)(100,0){4}{5}
%\Line[arrow,arrowinset=0.3,arrowaspect=1,arrowwidth=1,arrowpos=1,
%arrowstroke=3](95,0)(100,0)
\Text(25,110)[]{$2^+$}
\Text(25,10)[]{$1^-$}
\Text(110,110)[]{$3^+$}
\Text(110,80)[]{$4^+$}
\Text(110,50)[]{$5^+$}
\Text(110,20)[]{$6^-$}
\Text(110,-10)[]{$7^-$}
\Text(50,-20)[]{{\red{\bf 4.a)}}}

%\hspace*{1cm}

\SetWidth{1.}
\SetScale{0.7}
\ArrowLine(130,120)(180,120)
\Text(180,120)[]{{\blue{\large $\bullet$}}}
\Text(180,90)[]{{\blue{\large $\bullet$}}}
\Text(180,60)[]{{\blue{\large $\bullet$}}}
\Text(180,30)[]{{\blue{\large $\bullet$}}}
\Text(180,0)[]{{\blue{\large $\bullet$}}}
\ArrowLine(180,120)(180,90)
\ArrowLine(180,90)(180,60)
\ArrowLine(180,60)(180,30)
\ArrowLine(180,30)(180,0)
\ArrowLine(180,0)(130,0)
%\ArrowLine(100,25)(130,50)
\Photon(180,120)(230,120){4}{5}
%\Line[arrow,arrowinset=0.3,arrowaspect=1,arrowwidth=1,arrowpos=1,
%arrowstroke=3](95,90)(100,90)
\Photon(180,90)(230,90){4}{5}
%\Line[arrow,arrowinset=0.3,arrowaspect=1,arrowwidth=1,arrowpos=1,
%arrowstroke=3](95,60)(100,60)
\Photon(180,60)(230,60){4}{5}
%\Line[arrow,arrowinset=0.3,arrowaspect=1,arrowwidth=1,arrowpos=1,
%arrowstroke=3](95,30)(100,30)
\Photon(180,30)(230,30){4}{5}
\Photon(180,0)(230,0){4}{5}
%\Line[arrow,arrowinset=0.3,arrowaspect=1,arrowwidth=1,arrowpos=1,
%arrowstroke=3](95,0)(100,0)
\Text(155,110)[]{$2^+$}
\Text(155,10)[]{$1^-$}
\Text(240,110)[]{$3^+$}
\Text(240,80)[]{$4^+$}
\Text(240,50)[]{$6^-$}
\Text(240,20)[]{$5^+$}
\Text(240,-10)[]{$7^-$}
\Text(180,-20)[]{{\red{\bf 4.b)}}}

\SetWidth{1.}
\SetScale{0.7}
\ArrowLine(260,120)(310,120)
\Text(310,120)[]{{\blue{\large $\bullet$}}}
\Text(310,90)[]{{\blue{\large $\bullet$}}}
\Text(310,60)[]{{\blue{\large $\bullet$}}}
\Text(310,30)[]{{\blue{\large $\bullet$}}}
\Text(310,0)[]{{\blue{\large $\bullet$}}}
\ArrowLine(310,120)(310,90)
\ArrowLine(310,90)(310,60)
\ArrowLine(310,60)(310,30)
\ArrowLine(310,30)(310,0)
\ArrowLine(310,0)(260,0)
%\ArrowLine(100,25)(130,50)
\Photon(310,120)(360,120){4}{5}
%\Line[arrow,arrowinset=0.3,arrowaspect=1,arrowwidth=1,arrowpos=1,
%arrowstroke=3](95,90)(100,90)
\Photon(310,90)(360,90){4}{5}
%\Line[arrow,arrowinset=0.3,arrowaspect=1,arrowwidth=1,arrowpos=1,
%arrowstroke=3](95,60)(100,60)
\Photon(310,60)(360,60){4}{5}
%\Line[arrow,arrowinset=0.3,arrowaspect=1,arrowwidth=1,arrowpos=1,
%arrowstroke=3](95,30)(100,30)
\Photon(310,30)(360,30){4}{5}
\Photon(310,0)(360,0){4}{5}
%\Line[arrow,arrowinset=0.3,arrowaspect=1,arrowwidth=1,arrowpos=1,
%arrowstroke=3](95,0)(100,0)
\Text(285,110)[]{$2^+$}
\Text(285,10)[]{$1^-$}
\Text(370,110)[]{$3^+$}
\Text(370,80)[]{$6^-$}
\Text(370,50)[]{$4^+$}
\Text(370,20)[]{$5^+$}
\Text(370,-10)[]{$7^-$}
\Text(310,-20)[]{{\red{\bf 4.c)}}}

\end{picture}
\end{center}
\vspace*{0.5cm}
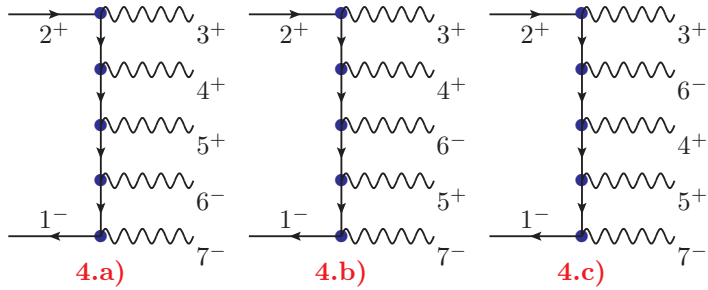
\centerline{\it Figure 4: Three types of non-vanishing Feynman Diagrams contributing to the term $A^\tree_7$ are shown. \label{figure4}} 
 
To realize the boson exchanging symmetry, we have used permutation group symbols. For example, in $S^2_7$, $[P(6,7)]$ means two terms should be added. In $S^3_7$, $[P(4,5)] \frac{\spb4.7 [5 | 2+3| 6\rangle}{S_{147}}$ means $ \frac{\spb5.7 [4 | 2+3| 6\rangle}{S_{157}} +  \frac{\spb4.7 [5 | 2+3| 6\rangle}{S_{147}}$. It is obvious boson exchanging symmetries among photons with the same helicities for $A^\tree_7$ are explicit. 

   \begin{center}
 \vspace*{-2.0cm}
\hspace*{2.0cm}
\begin{picture}(480,150)(0,0)
\SetWidth{1.}
\SetScale{0.6}
\ArrowLine(0,150)(50,150)
\Text(50,150)[]{{\blue{\large $\bullet$}}}
\Text(50,120)[]{{\blue{\large $\bullet$}}}
\Text(50,90)[]{{\blue{\large $\bullet$}}}
\Text(50,60)[]{{\blue{\large $\bullet$}}}
\Text(50,30)[]{{\blue{\large $\bullet$}}}
\Text(50,0)[]{{\blue{\large $\bullet$}}}
\ArrowLine(50,150)(50,120)
\ArrowLine(50,120)(50,90)
\ArrowLine(50,90)(50,60)
\ArrowLine(50,60)(50,30)
\ArrowLine(50,30)(50,0)
\ArrowLine(50,0)(0,0)
%\ArrowLine(100,25)(130,50)
\Photon(50,150)(100,150){4}{5}
\Photon(50,120)(100,120){4}{5}
%\Line[arrow,arrowinset=0.3,arrowaspect=1,arrowwidth=1,arrowpos=1,
%arrowstroke=3](95,90)(100,90)
\Photon(50,90)(100,90){4}{5}
%\Line[arrow,arrowinset=0.3,arrowaspect=1,arrowwidth=1,arrowpos=1,
%arrowstroke=3](95,60)(100,60)
\Photon(50,60)(100,60){4}{5}
%\Line[arrow,arrowinset=0.3,arrowaspect=1,arrowwidth=1,arrowpos=1,
%arrowstroke=3](95,30)(100,30)
\Photon(50,30)(100,30){4}{5}
\Photon(50,0)(100,0){4}{5}
%\Line[arrow,arrowinset=0.3,arrowaspect=1,arrowwidth=1,arrowpos=1,
%arrowstroke=3](95,0)(100,0)
\Text(25,140)[]{$2^+$}
\Text(25,10)[]{$1^-$}
\Text(110,140)[]{$3^+$}
\Text(110,110)[]{$4^+$}
\Text(110,80)[]{$5^+$}
\Text(110,50)[]{$6^+$}
\Text(110,20)[]{$7^-$}
\Text(110,-10)[]{$8^-$}
\Text(50,-20)[]{{\red{\bf 5.a)}}}

%\hspace*{1cm}

\SetWidth{1.}
\SetScale{0.6}
\ArrowLine(130,150)(180,150)
\Text(180,150)[]{{\blue{\large $\bullet$}}}
\Text(180,120)[]{{\blue{\large $\bullet$}}}
\Text(180,90)[]{{\blue{\large $\bullet$}}}
\Text(180,60)[]{{\blue{\large $\bullet$}}}
\Text(180,30)[]{{\blue{\large $\bullet$}}}
\Text(180,0)[]{{\blue{\large $\bullet$}}}
\ArrowLine(180,150)(180,120)
\ArrowLine(180,120)(180,90)
\ArrowLine(180,90)(180,60)
\ArrowLine(180,60)(180,30)
\ArrowLine(180,30)(180,0)
\ArrowLine(180,0)(130,0)
\Photon(180,150)(230,150){4}{5}
%\ArrowLine(100,25)(130,50)
\Photon(180,120)(230,120){4}{5}
%\Line[arrow,arrowinset=0.3,arrowaspect=1,arrowwidth=1,arrowpos=1,
%arrowstroke=3](95,90)(100,90)
\Photon(180,90)(230,90){4}{5}
%\Line[arrow,arrowinset=0.3,arrowaspect=1,arrowwidth=1,arrowpos=1,
%arrowstroke=3](95,60)(100,60)
\Photon(180,60)(230,60){4}{5}
%\Line[arrow,arrowinset=0.3,arrowaspect=1,arrowwidth=1,arrowpos=1,
%arrowstroke=3](95,30)(100,30)
\Photon(180,30)(230,30){4}{5}
\Photon(180,0)(230,0){4}{5}
%\Line[arrow,arrowinset=0.3,arrowaspect=1,arrowwidth=1,arrowpos=1,
%arrowstroke=3](95,0)(100,0)
\Text(155,140)[]{$2^+$}
\Text(240,140)[]{$3^+$}
\Text(155,10)[]{$1^-$}
\Text(240,110)[]{$4^+$}
\Text(240,80)[]{$5^+$}
\Text(240,50)[]{$7^-$}
\Text(240,20)[]{$6^+$}
\Text(240,-10)[]{$8^-$}
\Text(180,-20)[]{{\red{\bf 5.b)}}}

\SetWidth{1.}
\SetScale{0.6}
\ArrowLine(260,150)(310,150)
\Text(310,150)[]{{\blue{\large $\bullet$}}}
\Text(310,120)[]{{\blue{\large $\bullet$}}}
\Text(310,90)[]{{\blue{\large $\bullet$}}}
\Text(310,60)[]{{\blue{\large $\bullet$}}}
\Text(310,30)[]{{\blue{\large $\bullet$}}}
\Text(310,0)[]{{\blue{\large $\bullet$}}}
\ArrowLine(310,150)(310,120)
\ArrowLine(310,120)(310,90)
\ArrowLine(310,90)(310,60)
\ArrowLine(310,60)(310,30)
\ArrowLine(310,30)(310,0)
\ArrowLine(310,0)(260,0)
%\ArrowLine(100,25)(130,50)
\Photon(310,150)(360,150){4}{5}
\Photon(310,120)(360,120){4}{5}
%\Line[arrow,arrowinset=0.3,arrowaspect=1,arrowwidth=1,arrowpos=1,
%arrowstroke=3](95,90)(100,90)
\Photon(310,90)(360,90){4}{5}
%\Line[arrow,arrowinset=0.3,arrowaspect=1,arrowwidth=1,arrowpos=1,
%arrowstroke=3](95,60)(100,60)
\Photon(310,60)(360,60){4}{5}
%\Line[arrow,arrowinset=0.3,arrowaspect=1,arrowwidth=1,arrowpos=1,
%arrowstroke=3](95,30)(100,30)
\Photon(310,30)(360,30){4}{5}
\Photon(310,0)(360,0){4}{5}
%\Line[arrow,arrowinset=0.3,arrowaspect=1,arrowwidth=1,arrowpos=1,
%arrowstroke=3](95,0)(100,0)
\Text(285,140)[]{$2^+$}
\Text(285,10)[]{$1^-$}
\Text(370,140)[]{$3^+$}
\Text(370,110)[]{$4^+$}
\Text(370,80)[]{$7^-$}
\Text(370,50)[]{$5^+$}
\Text(370,20)[]{$6^+$}
\Text(370,-10)[]{$8^-$}
\Text(310,-20)[]{{\red{\bf 5.c)}}}

\SetWidth{1.}
\SetScale{0.6}
\ArrowLine(390,150)(440,150)
\Text(440,150)[]{{\blue{\large $\bullet$}}}
\Text(440,120)[]{{\blue{\large $\bullet$}}}
\Text(440,90)[]{{\blue{\large $\bullet$}}}
\Text(440,60)[]{{\blue{\large $\bullet$}}}
\Text(440,30)[]{{\blue{\large $\bullet$}}}
\Text(440,0)[]{{\blue{\large $\bullet$}}}
\ArrowLine(440,150)(440,120)
\ArrowLine(440,120)(440,90)
\ArrowLine(440,90)(440,60)
\ArrowLine(440,60)(440,30)
\ArrowLine(440,30)(440,0)
\ArrowLine(440,0)(390,0)
%\ArrowLine(100,25)(130,50)
\Photon(440,150)(490,150){4}{5}
\Photon(440,120)(490,120){4}{5}
%\Line[arrow,arrowinset=0.3,arrowaspect=1,arrowwidth=1,arrowpos=1,
%arrowstroke=3](95,90)(100,90)
\Photon(440,90)(490,90){4}{5}
%\Line[arrow,arrowinset=0.3,arrowaspect=1,arrowwidth=1,arrowpos=1,
%arrowstroke=3](95,60)(100,60)
\Photon(440,60)(490,60){4}{5}
%\Line[arrow,arrowinset=0.3,arrowaspect=1,arrowwidth=1,arrowpos=1,
%arrowstroke=3](95,30)(100,30)
\Photon(440,30)(490,30){4}{5}
\Photon(440,0)(490,0){4}{5}
%\Line[arrow,arrowinset=0.3,arrowaspect=1,arrowwidth=1,arrowpos=1,
%arrowstroke=3](95,0)(100,0)
\Text(415,140)[]{$2^+$}
\Text(415,10)[]{$1^-$}
\Text(500,140)[]{$3^+$}
\Text(500,110)[]{$7^-$}
\Text(500,80)[]{$4^+$}
\Text(500,50)[]{$5^+$}
\Text(500,20)[]{$6^+$}
\Text(500,-10)[]{$8^-$}
\Text(440,-20)[]{{\red{\bf 5.d)}}}

\end{picture}
\end{center}
\vspace*{0.5cm}
\centerline{\it Figure 5: Four types of non-vanishing Feynman Diagrams contributing to the term $A^\tree_8$ are shown. \label{figure5}} 

Similarly, the NMHV amplitude of $A^\tree_8(1_e^-,2_{\bar{e}}^+,3_{\gamma}^+,4_{\gamma}^+,5_{\gamma}^+, 6_{\gamma}^+, 7_{\gamma}^-,8_\gamma^-)$ can be computed efficiently by the off-shell amplitude method. We can use the Feynman diagrams given in Figure 5 as a guide to compute the results of each helicity configuration by using the off-shell amplitudes and permutation symmetries. The results are organized into four helicity configurations as $+ ++++-- -$ (denoted as $S^1_{8N}$), $+ +++-+- -$ (denoted as $S^2_{8N}$), $+ ++-++- -$ (denoted as $S^3_{8N}$), and $+ +-+++- -$ (denoted as $S^4_{8N}$). The total amplitude can be organized into the following form
\bea
A^\tree_{8N}(1_e^-,2_{\bar{e}}^+,3_{\gamma}^+,4_{\gamma}^+,5_{\gamma}^+, 6_{\gamma}^+, 7_{\gamma}^-, 8_\gamma^-) &=& Q_e^6 \left \{  S^1_{8N} + S^2_{8N} + S_{8N}^3 + S_{8N}^4 \right \} \label{nmhv8}\,, 
\eea
 where each $S^i_8$ term corresponding to each of figures in Figure 5 is given as
 \bea
 S_{8N}^1 &=& \frac{\spa1.2^3 \spb1.2 \,\, \langle 1 | 7 + 8 | 2] \,\, S_{178} }{\prod_{i=3,4,5,6} \spa1.i \spa2.i  \prod_{j=7,8} \spb1.j \spb2.j  }\,,\\
 S_{8N}^2&=&  \frac{\spa1.2^2 }{\prod_{i=3,4,5,6} \spa1.i } [P(345,6)] [P(7,8)] \left\{ \frac{\langle 1 | 3 + 4 + 5 | 2 ]  }{\prod_{j=3,4,5}\spa2.j }  \frac{\spa1.8}{\spb1.8} \frac{[6 | 1+8 | 7 \rangle }{\spb2.7 S_{168}} \right \}\, ,\label{a8n2}\\
 S_{8N}^3 & = & \frac{\spa1.2}{\prod_{i=3,4,5,6} \spa1.i} [P(7,8)] \frac{\spa1.8^2}{\spb2.7 \spb1.8 } \left \{ [P(34,56)]  \right. \nonumber \\
 && \left. \frac{\langle 1 | 3+4 | 2] }{\spa2.3 \spa2.4 S_{1568}}  [P(5,6)] \frac{\spb8.6 [5 | 1 + 6+8 | 7\rangle}{S_{168}}  \right\} \label{a8n3} \,,\\
S^4_{8N} & = & 
\spa1.2 [P(7,8)]  \frac{\spa1.8^2}{\spb2.7 \spb1.8 }  \left \{ [P(34,56)]  \right. \nonumber \\
 &&  \left.  \frac{1}{\spa1.5 \spa1.6 S_{1568} } \left (  [P(3,4)]\frac{\spb2.3}{\spa2.3}\frac{[4 | 2 + 3 | 7 \rangle }{\spa1.4 S_{237}} \right )  \left (  [P(5,6)] \frac{\spb8.6 [5 | 1 + 6+8 | 7\rangle}{S_{168}}  \right )  \right \} \label{a8n4}\,,
 \eea
where $[P(345,6)]=E + P_{36} + P_{46} + P_{56}$ which reflects the exchanging symmetry among $3$, $4$, and $5$. And $[P(34, 56)] = E + P_{35} + P_{36} + P_{45} + P_{46} + P_{35} P_{46}$, which express the exchanging symmetry between $3$ and $4$, and the symmetry between $5$ and $6$. Obviously, according to these convention, there exists a relation that $[P(3,4,5,6)] = [P(34,56)] [P(3,4)] [P(5,6)]$.
 \begin{center}
 \vspace*{-2.0cm}
\hspace*{1.0cm}
\begin{picture}(480,150)(0,0)
\SetWidth{1.}
\SetScale{0.5}
\ArrowLine(0,150)(50,150)
\Text(50,150)[]{{\blue{\large $\bullet$}}}
\Text(50,120)[]{{\blue{\large $\bullet$}}}
\Text(50,90)[]{{\blue{\large $\bullet$}}}
\Text(50,60)[]{{\blue{\large $\bullet$}}}
\Text(50,30)[]{{\blue{\large $\bullet$}}}
\Text(50,0)[]{{\blue{\large $\bullet$}}}
\ArrowLine(50,150)(50,120)
\ArrowLine(50,120)(50,90)
\ArrowLine(50,90)(50,60)
\ArrowLine(50,60)(50,30)
\ArrowLine(50,30)(50,0)
\ArrowLine(50,0)(0,0)
%\ArrowLine(100,25)(130,50)
\Photon(50,150)(100,150){4}{5}
\Photon(50,120)(100,120){4}{5}
%\Line[arrow,arrowinset=0.3,arrowaspect=1,arrowwidth=1,arrowpos=1,
%arrowstroke=3](95,90)(100,90)
\Photon(50,90)(100,90){4}{5}
%\Line[arrow,arrowinset=0.3,arrowaspect=1,arrowwidth=1,arrowpos=1,
%arrowstroke=3](95,60)(100,60)
\Photon(50,60)(100,60){4}{5}
%\Line[arrow,arrowinset=0.3,arrowaspect=1,arrowwidth=1,arrowpos=1,
%arrowstroke=3](95,30)(100,30)
\Photon(50,30)(100,30){4}{5}
\Photon(50,0)(100,0){4}{5}
%\Line[arrow,arrowinset=0.3,arrowaspect=1,arrowwidth=1,arrowpos=1,
%arrowstroke=3](95,0)(100,0)
\Text(25,140)[]{$2^+$}
\Text(25,10)[]{$1^-$}
\Text(110,140)[]{$3^+$}
\Text(110,110)[]{$4^+$}
\Text(110,80)[]{$5^+$}
\Text(110,50)[]{$6^-$}
\Text(110,20)[]{$7^-$}
\Text(110,-10)[]{$8^-$}
\Text(50,-20)[]{{\red{\bf 6.a)}}}

%\hspace*{1cm}

\SetWidth{1.}
\SetScale{0.5}
\ArrowLine(130,150)(180,150)
\Text(180,150)[]{{\blue{\large $\bullet$}}}
\Text(180,120)[]{{\blue{\large $\bullet$}}}
\Text(180,90)[]{{\blue{\large $\bullet$}}}
\Text(180,60)[]{{\blue{\large $\bullet$}}}
\Text(180,30)[]{{\blue{\large $\bullet$}}}
\Text(180,0)[]{{\blue{\large $\bullet$}}}
\ArrowLine(180,150)(180,120)
\ArrowLine(180,120)(180,90)
\ArrowLine(180,90)(180,60)
\ArrowLine(180,60)(180,30)
\ArrowLine(180,30)(180,0)
\ArrowLine(180,0)(130,0)
\Photon(180,150)(230,150){4}{5}
%\ArrowLine(100,25)(130,50)
\Photon(180,120)(230,120){4}{5}
%\Line[arrow,arrowinset=0.3,arrowaspect=1,arrowwidth=1,arrowpos=1,
%arrowstroke=3](95,90)(100,90)
\Photon(180,90)(230,90){4}{5}
%\Line[arrow,arrowinset=0.3,arrowaspect=1,arrowwidth=1,arrowpos=1,
%arrowstroke=3](95,60)(100,60)
\Photon(180,60)(230,60){4}{5}
%\Line[arrow,arrowinset=0.3,arrowaspect=1,arrowwidth=1,arrowpos=1,
%arrowstroke=3](95,30)(100,30)
\Photon(180,30)(230,30){4}{5}
\Photon(180,0)(230,0){4}{5}
%\Line[arrow,arrowinset=0.3,arrowaspect=1,arrowwidth=1,arrowpos=1,
%arrowstroke=3](95,0)(100,0)
\Text(155,140)[]{$2^+$}
\Text(240,140)[]{$3^+$}
\Text(155,10)[]{$1^-$}
\Text(240,110)[]{$4^+$}
\Text(240,80)[]{$6^-$}
\Text(240,50)[]{$5^+$}
\Text(240,20)[]{$7^-$}
\Text(240,-10)[]{$8^-$}
\Text(180,-20)[]{{\red{\bf 6.b)}}}

\SetWidth{1.}
\SetScale{0.5}
\ArrowLine(260,150)(310,150)
\Text(310,150)[]{{\blue{\large $\bullet$}}}
\Text(310,120)[]{{\blue{\large $\bullet$}}}
\Text(310,90)[]{{\blue{\large $\bullet$}}}
\Text(310,60)[]{{\blue{\large $\bullet$}}}
\Text(310,30)[]{{\blue{\large $\bullet$}}}
\Text(310,0)[]{{\blue{\large $\bullet$}}}
\ArrowLine(310,150)(310,120)
\ArrowLine(310,120)(310,90)
\ArrowLine(310,90)(310,60)
\ArrowLine(310,60)(310,30)
\ArrowLine(310,30)(310,0)
\ArrowLine(310,0)(260,0)
%\ArrowLine(100,25)(130,50)
\Photon(310,150)(360,150){4}{5}
\Photon(310,120)(360,120){4}{5}
%\Line[arrow,arrowinset=0.3,arrowaspect=1,arrowwidth=1,arrowpos=1,
%arrowstroke=3](95,90)(100,90)
\Photon(310,90)(360,90){4}{5}
%\Line[arrow,arrowinset=0.3,arrowaspect=1,arrowwidth=1,arrowpos=1,
%arrowstroke=3](95,60)(100,60)
\Photon(310,60)(360,60){4}{5}
%\Line[arrow,arrowinset=0.3,arrowaspect=1,arrowwidth=1,arrowpos=1,
%arrowstroke=3](95,30)(100,30)
\Photon(310,30)(360,30){4}{5}
\Photon(310,0)(360,0){4}{5}
%\Line[arrow,arrowinset=0.3,arrowaspect=1,arrowwidth=1,arrowpos=1,
%arrowstroke=3](95,0)(100,0)
\Text(285,140)[]{$2^+$}
\Text(285,10)[]{$1^-$}
\Text(370,140)[]{$3^+$}
\Text(370,110)[]{$6^-$}
\Text(370,80)[]{$4^+$}
\Text(370,50)[]{$5^+$}
\Text(370,20)[]{$7^-$}
\Text(370,-10)[]{$8^-$}
\Text(310,-20)[]{{\red{\bf 6.c)}}}

\SetWidth{1.}
\SetScale{0.5}
\ArrowLine(390,150)(440,150)
\Text(440,150)[]{{\blue{\large $\bullet$}}}
\Text(440,120)[]{{\blue{\large $\bullet$}}}
\Text(440,90)[]{{\blue{\large $\bullet$}}}
\Text(440,60)[]{{\blue{\large $\bullet$}}}
\Text(440,30)[]{{\blue{\large $\bullet$}}}
\Text(440,0)[]{{\blue{\large $\bullet$}}}
\ArrowLine(440,150)(440,120)
\ArrowLine(440,120)(440,90)
\ArrowLine(440,90)(440,60)
\ArrowLine(440,60)(440,30)
\ArrowLine(440,30)(440,0)
\ArrowLine(440,0)(390,0)
%\ArrowLine(100,25)(130,50)
\Photon(440,150)(490,150){4}{5}
\Photon(440,120)(490,120){4}{5}
%\Line[arrow,arrowinset=0.3,arrowaspect=1,arrowwidth=1,arrowpos=1,
%arrowstroke=3](95,90)(100,90)
\Photon(440,90)(490,90){4}{5}
%\Line[arrow,arrowinset=0.3,arrowaspect=1,arrowwidth=1,arrowpos=1,
%arrowstroke=3](95,60)(100,60)
\Photon(440,60)(490,60){4}{5}
%\Line[arrow,arrowinset=0.3,arrowaspect=1,arrowwidth=1,arrowpos=1,
%arrowstroke=3](95,30)(100,30)
\Photon(440,30)(490,30){4}{5}
\Photon(440,0)(490,0){4}{5}
%\Line[arrow,arrowinset=0.3,arrowaspect=1,arrowwidth=1,arrowpos=1,
%arrowstroke=3](95,0)(100,0)
\Text(415,140)[]{$2^+$}
\Text(415,10)[]{$1^-$}
\Text(500,140)[]{$3^+$}
\Text(500,110)[]{$4^+$}
\Text(500,80)[]{$6^-$}
\Text(500,50)[]{$7^-$}
\Text(500,20)[]{$5^+$}
\Text(500,-10)[]{$8^-$}
\Text(440,-20)[]{{\red{\bf 6.d)}}}

\SetWidth{1.}
\SetScale{0.5}
\ArrowLine(520,150)(570,150)
\Text(570,150)[]{{\blue{\large $\bullet$}}}
\Text(570,120)[]{{\blue{\large $\bullet$}}}
\Text(570,90)[]{{\blue{\large $\bullet$}}}
\Text(570,60)[]{{\blue{\large $\bullet$}}}
\Text(570,30)[]{{\blue{\large $\bullet$}}}
\Text(570,0)[]{{\blue{\large $\bullet$}}}
\ArrowLine(570,150)(570,120)
\ArrowLine(570,120)(570,90)
\ArrowLine(570,90)(570,60)
\ArrowLine(570,60)(570,30)
\ArrowLine(570,30)(570,0)
\ArrowLine(570,0)(520,0)
%\ArrowLine(100,25)(130,50)
\Photon(570,150)(620,150){4}{5}
\Photon(570,120)(620,120){4}{5}
%\Line[arrow,arrowinset=0.3,arrowaspect=1,arrowwidth=1,arrowpos=1,
%arrowstroke=3](95,90)(100,90)
\Photon(570,90)(620,90){4}{5}
%\Line[arrow,arrowinset=0.3,arrowaspect=1,arrowwidth=1,arrowpos=1,
%arrowstroke=3](95,60)(100,60)
\Photon(570,60)(620,60){4}{5}
%\Line[arrow,arrowinset=0.3,arrowaspect=1,arrowwidth=1,arrowpos=1,
%arrowstroke=3](95,30)(100,30)
\Photon(570,30)(620,30){4}{5}
\Photon(570,0)(620,0){4}{5}
%\Line[arrow,arrowinset=0.3,arrowaspect=1,arrowwidth=1,arrowpos=1,
%arrowstroke=3](95,0)(100,0)
\Text(545,140)[]{$2^+$}
\Text(545,10)[]{$1^-$}
\Text(630,140)[]{$3^+$}
\Text(630,110)[]{$6^-$}
\Text(630,80)[]{$4^+$}
\Text(630,50)[]{$7^-$}
\Text(630,20)[]{$5^+$}
\Text(630,-10)[]{$8^-$}
\Text(570,-20)[]{{\red{\bf 6.e)}}}

\SetWidth{1.}
\SetScale{0.5}
\ArrowLine(650,150)(700,150)
\Text(700,150)[]{{\blue{\large $\bullet$}}}
\Text(700,120)[]{{\blue{\large $\bullet$}}}
\Text(700,90)[]{{\blue{\large $\bullet$}}}
\Text(700,60)[]{{\blue{\large $\bullet$}}}
\Text(700,30)[]{{\blue{\large $\bullet$}}}
\Text(700,0)[]{{\blue{\large $\bullet$}}}
\ArrowLine(700,150)(700,120)
\ArrowLine(700,120)(700,90)
\ArrowLine(700,90)(700,60)
\ArrowLine(700,60)(700,30)
\ArrowLine(700,30)(700,0)
\ArrowLine(700,0)(650,0)
%\ArrowLine(100,25)(130,50)
\Photon(700,150)(750,150){4}{5}
\Photon(700,120)(750,120){4}{5}
%\Line[arrow,arrowinset=0.3,arrowaspect=1,arrowwidth=1,arrowpos=1,
%arrowstroke=3](95,90)(100,90)
\Photon(700,90)(750,90){4}{5}
%\Line[arrow,arrowinset=0.3,arrowaspect=1,arrowwidth=1,arrowpos=1,
%arrowstroke=3](95,60)(100,60)
\Photon(700,60)(750,60){4}{5}
%\Line[arrow,arrowinset=0.3,arrowaspect=1,arrowwidth=1,arrowpos=1,
%arrowstroke=3](95,30)(100,30)
\Photon(700,30)(750,30){4}{5}
\Photon(700,0)(750,0){4}{5}
%\Line[arrow,arrowinset=0.3,arrowaspect=1,arrowwidth=1,arrowpos=1,
%arrowstroke=3](95,0)(100,0)
\Text(675,140)[]{$2^+$}
\Text(675,10)[]{$1^-$}
\Text(760,140)[]{$3^+$}
\Text(760,110)[]{$6^-$}
\Text(760,80)[]{$7^-$}
\Text(760,50)[]{$4^+$}
\Text(760,20)[]{$5^+$}
\Text(760,-10)[]{$8^-$}
\Text(700,-20)[]{{\red{\bf 6.f)}}}

\end{picture}
\end{center}
\vspace*{0.5cm}
\centerline{\it Figure 6: Four types of non-vanishing Feynman Diagrams contributing to the term $A^\tree_8$ are shown. \label{figure6}}

The N$^2$MHV amplitude of $A^\tree_8(1_e^-,2_{\bar{e}}^+,3_{\gamma}^+,4_{\gamma}^+,5_{\gamma}^+, 6_{\gamma}^-, 7_{\gamma}^-,8_\gamma^-)$ can be computed from the Feynman diagrams given in Figure 6, and can be expressed as
\bea
A^\tree_{8N^2}(1_e^-,2_{\bar{e}}^+,3_{\gamma}^+,4_{\gamma}^+,5_{\gamma}^+, 6_{\gamma}^+, 7_{\gamma}^-, 8_\gamma^-) &=& Q_e^6 \left \{  S^1_{8N^2} + S^2_{8N^2} + S_{8N^2}^3 +  S^4_{8N^2} + S^5_{8N^2} + S_{8N^2}^6 \right \} \label{nmhv8n2}\,, 
\eea
where each of $S^i_{8N^2}$ corresponding to each of diagrams in Figure 6 is given below 
\bea
 S^1_{8N^2} & = &  \frac{\spa1.2^2 \spb1.2^2 \,\, \langle 1 | 6 + 7 + 8 | 2] \,\, S_{1678} }{\prod_{i=3,4,5} \spa1.i \spa2.i  \prod_{j=6,7,8} \spb1.j \spb2.j  }\,,\\
 S^2_{8N^2} & = & [P(6,78)] [P(34,5)]  \nonumber \\
 && \left[ \frac{\spa1.2 \spb1.2}{(\prod_{i=3,4} \spa1.i \spa2.i) (\prod_{j=7,8}) \spb1.j \spb2.j} \frac{\langle 1 | 7+8 | 2] \langle 1 | 3 + 4 \ 2] [5 | 1 + 7 + 8 | 6 \rangle}{\spa1.5 \spb2.6 S_{1578}}  \right ]\,, \label{a8nn2}\\
S^3_{8N^2} &=&[P(34,5)]  \left\{ [P(3,4)]  \frac{\spb2.3}{\spa2.3 \spa1.4}  \left[ [P(6,78)] \right.\right. \nonumber \\
&& \left.\left. \frac{\spb1.2}{(\prod_{j=7,8} \spb1.j \spb2.j)} \frac{[2 | 7+8 | 1 \rangle}{\spa1.5 \spb2.6} \frac{[4 | 2 + 3 | 6 \rangle [5 | 7 + 8| 1\rangle}{S_{236} S_{1578} }  \right] \right \} \,,\\
 S^4_{8N^2} &=& [P(6,78)]\left\{  [P(7,8)]  \frac{\spa1.8}{\spb2.7 \spb1.8}  \left[  [P(34,5)] \right.\right. \nonumber \\
&&\left. \left.  \frac{\spa1.2}{(\prod_{j=3,4} \spa1.j \spa2.j)} \frac{[2 | 3 +4 | 1 \rangle}{\spa1.5 \spb2.6} \frac{[5 | 1+8 | 7 \rangle [2 | 3+4| 6\rangle}{S_{158} S_{1578} }  \right ] \right\} \,,\\
S^5_{8N^2} & =& [P(34,5)] [P(6,78)] \left \{ \frac{1}{\spa1.5 \spb2.6} \left [ [P(7,8)]  [P(3,4)] \left (  \right.\right. \right. \nonumber \\
&& \left.\left.\left. \frac{\spb2.3}{\spa2.3 \spa1.4} \frac{\spa1.8}{\spb2.7 \spb1.8} \frac{[4 | 2+3| 6 \rangle [5 | 1 + 8| 7 \rangle [2| 5 + 7 + 8 | 1 \rangle}{S_{236} S_{158} S_{1578} } \right) \right ]  \right \}  \label{a8nn5} \,,\\
S^6_{8N^2} & = &  [P(3,45)] [P(67,8)]  \left \{  \frac{\spb2.3^2 \spa1.8^2}{\spa2.3\spa1.4\spa1.5 \spb2.6 \spb2.7 \spb1.8 S_{1458} } \left( \right.\right.  \nonumber  \\
&& \left. \left. [P(4,5)] [P(6,7)] \frac{\spa3.6 \spb5.8 [4 | 2 + 3 + 6 | 7 \rangle}{S_{158} S_{236} } \right)  \right \}  \,.\label{a8nn6}
\eea
It should be pointed out that by using the off-shell current method given in \cite{Berends:1987me} and permutation group for boson exchanging symmetries, we can put the amplitudes of a given helicity configuration in an elegant form with less number of terms when compared with the direct Feynman diagram method. These analytic form of amplitudes are helpful for us to understand the properties all shifts in the BCFW method. 

\section{Amplitudes of the process $0 \to e^- e^+ 4\gamma$  by the BCFW method}\label{bcfwshifts}
To compute the amplitudes of the process $0 \to e^- e^+ 4\gamma$ by the BCFW method, we can organize the diagrams with three types of topologies, i.e. $ 3 \otimes 5$ (denoted as $P$ diagrams as shown in Figure 7, where $3$ means the left side has 3-point amplitudes and the right side has 5-point amplitudes), $ 4 \otimes 4$ (denoted as $R$ diagrams as shown in Figure 8 where both left and right sides have 4-point amplitudes), and $5 \otimes 3$ (denoted as $Q$ diagrams as shown in Figure 9 where the left side has 5-point amplitudes and the right side has 3-point amplitudes). Since the shift $[1,2\rangle$ is a natural choice to factorize the total amplitude due to the charge conservation law, and there are 12 diagrams in total. Below we describe all these diagrams in details. The diagrams for other shifts are just subsets of these 12 diagrams.

The topology $ 3 \otimes 5$ includes 4 diagrams given in Figure 7, which is labelled as $P_1$, $P_2$, $P_3$, and $P_4$. In these four diagrams, there exists an exchange symmetry between $P_1$ and $P_3$ with $3 \leftrightarrow 4 $, and between $P_2$ and $P_4$ with $5 \leftrightarrow 6$. Such exchanging symmetries can be used to simplify the calculation procedure. The momentum of the internal lines are denotes as $P_1$, $P_2$, $P_3$, and $P_4$, respectively. For example, $P_1= \hat{k}_1 + k_3 = - (\hat{k}_2 + k_4 + k_5 + k_6) $. The arrows in the fermion lines denote the direction of particle flow. 
 \begin{center}
 \vspace*{-2.0cm}
\hspace*{1.5cm}
\begin{picture}(440,120)(0,0)
\SetWidth{1.}
\SetScale{0.7}
%\ArrowLine(70,45)(20,90)
\ArrowLine(0,50)(0,0)
\Photon(0,50)(0,100){4}{5}
\ECirc(0,50){10}
\ArrowLine(30,50)(0,50)
\ArrowLine(60,50)(30,50)
\ECirc(60,50){10}
\ArrowLine(60,0)(60,50)
\Photon(60,50)(60,100){4}{5}
%\ArrowLine(120,45)(170,90)
\Photon(60,50)(100,80){4}{5}
%\ArrowLine(120,45)(170,60)
\Photon(60,50)(100,30){4}{5}
%\ArrowLine(120,45)(170,30)
\Text(-10,25)[]{$1^-$}
\Text(50,10)[]{$2^+$}
\Text(-10,80)[]{$3^+$}
\Text(75,90)[]{$4^+$}
\Text(100,60)[]{$5^-$}
\Text(90,20)[]{$6^-$}
%\Text(110,100)[]{$5^-$}
%\Text(120,10)[]{$3^+$}
%\hspace*{1cm}
\Text(30,-15)[]{{\red{\bf $P_1$ }}}

\SetWidth{1.}
\SetScale{0.7}
%\ArrowLine(70,45)(20,90)
\ArrowLine(135,50)(135,0)
\Photon(135,50)(135,100){4}{5}
\ECirc(135,50){10}
\ArrowLine(165,50)(135,50)
\ArrowLine(195,50)(165,50)
\ECirc(195,50){10}
\ArrowLine(195,0)(195,50)
\Photon(195,50)(195,100){4}{5}
%\ArrowLine(120,45)(170,90)
\Photon(195,50)(235,80){4}{5}
%\ArrowLine(120,45)(170,60)
\Photon(195,50)(235,30){4}{5}
\Text(125,25)[]{$1^-$}
\Text(185,10)[]{$2^+$}
\Text(125,80)[]{$5^-$}
\Text(210,90)[]{$3^+$}
\Text(235,60)[]{$4^+$}
\Text(225,20)[]{$6^-$}
%\ArrowLine(120,45)(170,30)
%\Text(20,100)[]{$2^-$}
%\Text(20,25)[]{$1^+$}
%\Text(50,130)[]{$4^+$}
%\Text(80,130)[]{$6^-$}
%\Text(110,100)[]{$5^-$}
%\Text(120,10)[]{$3^+$}
%\hspace*{1cm}
\Text(165,-15)[]{{\red{\bf $P_2$ }}}

\SetWidth{1.}
\SetScale{0.7}
%\ArrowLine(70,45)(20,90)
\ArrowLine(270,50)(270,0)
\Photon(270,50)(270,100){4}{5}
\ECirc(270,50){10}
\ArrowLine(300,50)(270,50)
\ArrowLine(330,50)(300,50)
\ECirc(330,50){10}
\ArrowLine(330,0)(330,50)
\Photon(330,50)(330,100){4}{5}
%\ArrowLine(120,45)(170,90)
\Photon(330,50)(370,80){4}{5}
%\ArrowLine(120,45)(170,60)
\Photon(330,50)(370,30){4}{5}
\Text(260,25)[]{$1^-$}
\Text(320,10)[]{$2^+$}
\Text(260,80)[]{$4^+$}
\Text(345,90)[]{$3^+$}
\Text(370,60)[]{$5^-$}
\Text(360,20)[]{$6^-$}
%\ArrowLine(120,45)(170,30)
%\Text(20,100)[]{$2^-$}
%\Text(20,25)[]{$1^+$}
%\Text(50,130)[]{$4^+$}
%\Text(80,130)[]{$6^-$}
%\Text(110,100)[]{$5^-$}
%\hspace*{1cm}
\Text(300,-15)[]{{\red{\bf $P_3$}}}

\SetWidth{1.}
\SetScale{0.7}
%\ArrowLine(70,45)(20,90)
\ArrowLine(405,50)(405,0)
\Photon(405,50)(405,100){4}{5}
\ECirc(405,50){10}
\ArrowLine(435,50)(405,50)
\ArrowLine(465,50)(435,50)
\ECirc(465,50){10}
\ArrowLine(465,0)(465,50)
\Photon(465,50)(465,100){4}{5}
%\ArrowLine(120,45)(170,90)
\Photon(465,50)(505,80){4}{5}
%\ArrowLine(120,45)(170,60)
\Photon(465,50)(505,30){4}{5}
\Text(395,25)[]{$1^-$}
\Text(455,10)[]{$2^+$}
\Text(395,80)[]{$6^-$}
\Text(480,90)[]{$3^+$}
\Text(505,60)[]{$4^+$}
\Text(495,20)[]{$5^-$}
%\ArrowLine(120,45)(170,30)
%\Text(20,100)[]{$2^-$}
%\Text(20,25)[]{$1^+$}
%\Text(50,130)[]{$4^+$}
%\Text(80,130)[]{$6^-$}
%\Text(110,100)[]{$5^-$}
%\Text(120,10)[]{$3^+$}
%\hspace*{1cm}
\Text(435,-15)[]{{\red{\bf $P_4$}}}

\end{picture}
\end{center}
\vspace*{0.5cm}
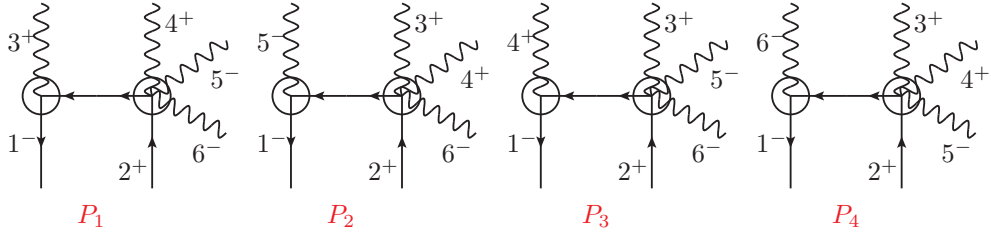
\centerline{\it Figure 7: The four diagrams for $3\otimes 5$ topology are shown in the $[1, 2 \rangle $ shift. } 

The topology $ 4 \otimes 4$ includes 4 diagrams given in Figure 8, which is labelled as $R_1$, $R_2$, $R_3$, and $R_4$. In these four diagrams, there exists an exchange symmetry between $R_1$ and $R_2$ with $3 \leftrightarrow 4 $, and between $R_1$ and $R_3$ with $5 \leftrightarrow 6$. Such exchanging symmetries can be used to simplify the calculation procedure. The momentum of the internal lines are denotes as $R_1$, $R_2$, $R_3$, and $R_4$, respectively. For example, $R_1= \hat{k}_1 + k_3 + k_5 = - ({\hat k}_2 + k_4 + k_6 )$. 

 \begin{center}
 \vspace*{-2.0cm}
\hspace*{1.5cm}
\begin{picture}(440,120)(0,0)
\SetWidth{1.}
\SetScale{0.7}
%\ArrowLine(70,45)(20,90)
\ArrowLine(0,50)(0,0)
\Photon(0,50)(0,100){4}{5}
\ECirc(0,50){10}
\ArrowLine(30,50)(0,50)
\ArrowLine(60,50)(10,50)
\ECirc(60,50){10}
\ArrowLine(60,0)(60,50)
\Photon(60,50)(60,100){4}{5}
%\ArrowLine(120,45)(170,90)
\Photon(-30,50)(0,50){4}{5}
%\ArrowLine(120,45)(170,60)
\Photon(60,50)(90,50){4}{5}
%\ArrowLine(120,45)(170,30)
\Text(-10,25)[]{$1^-$}
\Text(50,10)[]{$2^+$}
\Text(-35,40)[]{$3^+$}
\Text(80,90)[]{$4^+$}
\Text(-10,90)[]{$5^-$}
\Text(85,40)[]{$6^-$}
%\Text(110,100)[]{$5^-$}
%\Text(120,10)[]{$3^+$}
%\hspace*{1cm}
\Text(30,-15)[]{{\red{\bf $R_1$ }}}

\SetWidth{1.}
\SetScale{0.7}
%\ArrowLine(70,45)(20,90)
\ArrowLine(140,50)(140,0)
\Photon(140,50)(140,100){4}{5}
\ECirc(140,50){10}
\ArrowLine(170,50)(140,50)
\ArrowLine(200,50)(170,50)
\ECirc(200,50){10}
\ArrowLine(200,0)(200,50)
\Photon(200,50)(200,100){4}{5}
%\ArrowLine(120,45)(170,90)
\Photon(110,50)(140,50){4}{5}
%\ArrowLine(120,45)(170,60)
\Photon(200,50)(230,50){4}{5}
%\ArrowLine(120,45)(170,30)
\Text(130,25)[]{$1^-$}
\Text(190,10)[]{$2^+$}
\Text(110,40)[]{$4^+$}
\Text(220,90)[]{$3^+$}
\Text(130,90)[]{$5^-$}
\Text(225,40)[]{$6^-$}
%\Text(110,100)[]{$5^-$}
%\Text(120,10)[]{$3^+$}
\Text(170,-15)[]{{\red{\bf $R_2$ }}}

\SetWidth{1.}
\SetScale{0.7}
%\ArrowLine(70,45)(20,90)
\ArrowLine(280,50)(280,0)
\Photon(280,50)(280,100){4}{5}
\ECirc(280,50){10}
\ArrowLine(310,50)(280,50)
\ArrowLine(340,50)(310,50)
\ECirc(340,50){10}
\ArrowLine(340,0)(340,50)
\Photon(340,50)(340,100){4}{5}
%\ArrowLine(120,45)(170,90)
\Photon(250,50)(280,50){4}{5}
%\ArrowLine(120,45)(170,60)
\Photon(340,50)(370,50){4}{5}
%\ArrowLine(120,45)(170,30)
\Text(270,25)[]{$1^-$}
\Text(330,10)[]{$2^+$}
\Text(250,40)[]{$3^+$}
\Text(360,90)[]{$4^+$}
\Text(270,90)[]{$6^-$}
\Text(365,40)[]{$5^-$}
%\Text(110,100)[]{$5^-$}
%\Text(120,10)[]{$3^+$}
\Text(300,-15)[]{{\red{\bf $R_3$ }}}
\SetWidth{1.}
\SetScale{0.7}
%\ArrowLine(70,45)(20,90)
\ArrowLine(420,50)(420,0)
\Photon(420,50)(420,100){4}{5}
\ECirc(420,50){10}
\ArrowLine(450,50)(420,50)
\ArrowLine(480,50)(450,50)
\ECirc(480,50){10}
\ArrowLine(480,0)(480,50)
\Photon(480,50)(480,100){4}{5}
%\ArrowLine(120,45)(170,90)
\Photon(390,50)(420,50){4}{5}
%\ArrowLine(120,45)(170,60)
\Photon(480,50)(520,50){4}{5}
%\ArrowLine(120,45)(170,30)
\Text(410,25)[]{$1^-$}
\Text(470,10)[]{$2^+$}
\Text(390,40)[]{$4^+$}
\Text(500,90)[]{$3^+$}
\Text(410,90)[]{$6^-$}
\Text(505,40)[]{$5^-$}
\Text(440,-15)[]{{\red{\bf $R_4$ }}}
\end{picture}
\end{center}
\vspace*{2.0mm}
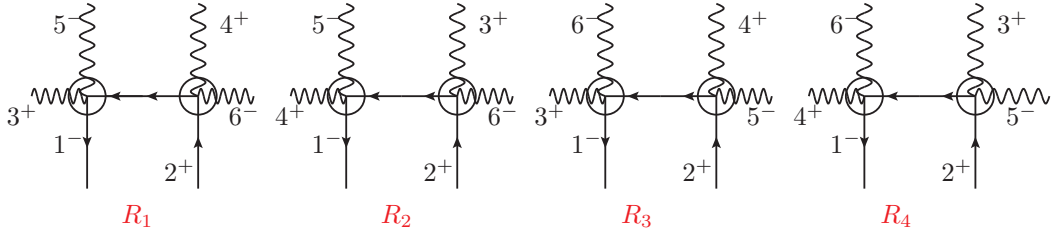
\centerline{\it Figure 8: The four diagrams for $4\otimes 4$ topology are shown in the $[1, 2 \rangle $ shift. } 

The topology $ 5 \otimes 3$ includes 4 diagrams given in Figure 9, which is labelled as $Q_1$, $Q_2$, $Q_3$, and $Q_4$. In these four diagrams, there exists an exchange symmetry between $Q_1$ and $Q_3$ with $3 \leftrightarrow 4 $, and between $Q_2$ and $Q_4$ with $5 \leftrightarrow 6$. Such exchanging symmetries can be used to simplify the calculation procedure. The momentum of the internal lines are denotes as $Q_1$, $Q_2$, $Q_3$, and $Q_4$, respectively. For example, $Q_1=\hat{k}_2 + k_3 = - (\hat{k}_1 + k_4 + k_5 + k_6)$.

 \begin{center}
 \vspace*{-2.0cm}
\hspace*{1.5cm}
\begin{picture}(440,120)(0,0)
\SetWidth{1.}
\SetScale{0.7}
%\ArrowLine(70,45)(20,90)
\ArrowLine(0,50)(0,0)
\Photon(0,50)(0,100){4}{5}
\ECirc(0,50){10}
\ArrowLine(30,50)(0,50)
\ArrowLine(60,50)(10,50)
\ECirc(60,50){10}
\ArrowLine(60,0)(60,50)
\Photon(60,50)(60,100){4}{5}
%\ArrowLine(120,45)(170,90)
\Photon(-30,75)(0,50){4}{5}
%\ArrowLine(120,45)(170,60)
\Photon(0,50)(-30,30){4}{5}
%\ArrowLine(120,45)(170,30)
\Text(-10,10)[]{$1^-$}
\Text(75,30)[]{$2^+$}
\Text(-35,35)[]{$6^-$}
\Text(75,90)[]{$3^+$}
\Text(-10,90)[]{$4^+$}
\Text(-35,70)[]{$5^-$}
%\Text(110,100)[]{$5^-$}
%\Text(120,10)[]{$3^+$}
%\hspace*{1cm}
\Text(30,-15)[]{{\red{\bf $Q_1$ }}}

\SetWidth{1.}
\SetScale{0.7}
%\ArrowLine(70,45)(20,90)
\ArrowLine(140,50)(140,0)
\Photon(140,50)(140,100){4}{5}
\ECirc(140,50){10}
\ArrowLine(170,50)(140,50)
\ArrowLine(200,50)(150,50)
\ECirc(200,50){10}
\ArrowLine(200,0)(200,50)
\Photon(200,50)(200,100){4}{5}
%\ArrowLine(120,45)(170,90)
\Photon(110,75)(140,50){4}{5}
%\ArrowLine(120,45)(170,60)
\Photon(140,50)(110,30){4}{5}
%\ArrowLine(120,45)(170,30)
\Text(130,10)[]{$1^-$}
\Text(215,30)[]{$2^+$}
\Text(105,35)[]{$6^-$}
\Text(215,90)[]{$5^-$}
\Text(130,90)[]{$3^+$}
\Text(105,70)[]{$4^+$}
%\Text(110,100)[]{$5^-$}
%\Text(120,10)[]{$3^+$}
%\hspace*{1cm}
\Text(170,-15)[]{{\red{\bf $Q_2$ }}}

\SetWidth{1.}
\SetScale{0.7}
%\ArrowLine(70,45)(20,90)
\ArrowLine(280,50)(280,0)
\Photon(280,50)(280,100){4}{5}
\ECirc(280,50){10}
\ArrowLine(310,50)(280,50)
\ArrowLine(340,50)(290,50)
\ECirc(340,50){10}
\ArrowLine(340,0)(340,50)
\Photon(340,50)(340,100){4}{5}
%\ArrowLine(120,45)(170,90)
\Photon(250,75)(280,50){4}{5}
%\ArrowLine(120,45)(170,60)
\Photon(280,50)(250,30){4}{5}
%\ArrowLine(120,45)(170,30)
\Text(270,10)[]{$1^-$}
\Text(355,30)[]{$2^+$}
\Text(245,35)[]{$6^-$}
\Text(355,90)[]{$4^+$}
\Text(270,90)[]{$3^+$}
\Text(245,70)[]{$5^-$}
%\Text(110,100)[]{$5^-$}
%\Text(120,10)[]{$3^+$}

\Text(310,-15)[]{{\red{\bf $Q_3$ }}}

\SetWidth{1.}
\SetScale{0.7}
%\ArrowLine(70,45)(20,90)
\ArrowLine(420,50)(420,0)
\Photon(420,50)(420,100){4}{5}
\ECirc(420,50){10}
\ArrowLine(450,50)(420,50)
\ArrowLine(490,50)(430,50)
\ECirc(480,50){10}
\ArrowLine(480,0)(480,50)
\Photon(480,50)(480,100){4}{5}
%\ArrowLine(120,45)(170,90)
\Photon(390,75)(420,50){4}{5}
%\ArrowLine(120,45)(170,60)
\Photon(420,50)(390,30){4}{5}
%\ArrowLine(120,45)(170,30)
\Text(410,10)[]{$1^-$}
\Text(495,30)[]{$2^+$}
\Text(385,35)[]{$5^-$}
\Text(495,90)[]{$6^-$}
\Text(410,90)[]{$3^+$}
\Text(385,70)[]{$4^+$}

\Text(450,-15)[]{{\red{\bf $Q_4$ }}}

\end{picture}
\end{center}
\vspace*{2.0mm}
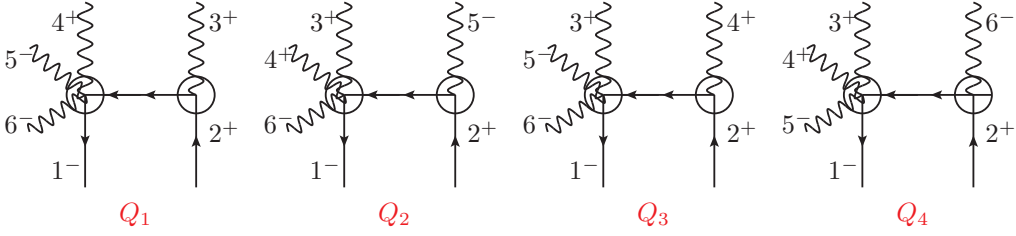
\centerline{\it Figure 9: The four diagrams for $5\otimes 3$ topology are shown in the $[1, 2 \rangle $ shift. } 

For the NMHV amplitude of the process $0 \to e^-(1^-) e^+(2^+)  \gamma^+(3^+) \gamma^+(4^+) \gamma^-(5^-) \gamma^-(6^-) $, there are 30 shifts that can be defined. From the result given in Eq. (\ref{nmhv}) and in terms of the highest $z$ power in the limit $z \to \infty$, they can be put into 5 categories: 
\begin{itemize}
\item 1) $z^2$: $[3, 5\rangle$ ($[4, 5\rangle$) and $[3, 6\rangle$ ($[4, 6\rangle$);
\item 2) $z^1$: $[3, 1\rangle$ ($[4, 1\rangle$) and $[2, 5\rangle$ ($[2, 6\rangle$);
\item 3) $z^0$: $[2,1\rangle$,  $[1,5\rangle$ ($[1,6\rangle$), and $[3,2\rangle$ ($[4,2\rangle$);
\item 4)$z^{-1}$: $[1, 3\rangle$ ($[1, 4\rangle$), $[5, 2\rangle$ ( $[6, 2\rangle$), $[5, 1\rangle$ ($[6, 1\rangle$), $[5, 6\rangle$ ($[6,5\rangle$), $[3, 4\rangle$ ($[4, 3\rangle$), $[2, 3\rangle$ ($[2, 4\rangle$);
\item 5)$z^{-2}$: $[1, 2 \rangle$, $[5,3\rangle$ ($[6,3\rangle$) and $[5,4\rangle$ ($[6,4\rangle$).
\end{itemize}

\begin{table}[htb]
\centering
\begin{tabular}{|c|c|c|c|c|c|}
\hline
%Amplitudes  &  & Form Gauge independent\\
\hline
shifts $[i, j\rangle$ $(i <j)$ & $[1, 2\rangle$ &  $[1, 3\rangle$ &  $[1, 4 \rangle$ &  $[1, 5\rangle$ &  $[1, 6\rangle$    \\
\hline
k &-2  & -2 & -2 & 0 & 0  \\
\hline
shifts $[i, j\rangle$ $(i <j)$ &  $[2, 3 \rangle$ & $[2, 4\rangle$ &  $[2, 5\rangle$ &  $[2, 6 \rangle$ &  $[3, 4\rangle$    \\
\hline
k & -1 & -1  & 1 & 1  & -1 \\
\hline
shifts $[i, j\rangle$ $(i <j)$ &  $[3, 5\rangle$ &  $[3, 6 \rangle$ & $[4, 5\rangle$ &  $[4, 6\rangle$ &  $[5, 6 \rangle$  \\
\hline
k & 2 & 2 & 2  & 2  & -1 \\
\hline
%Amplitudes  &  & Form Gauge independent\\
\hline
shifts $[i, j\rangle$ $(i > j)$ & $[ 2, 1\rangle$ &  $[3,1\rangle$ &  $[4, 1 \rangle$ &  $[5, 1\rangle$ &  $[6, 1\rangle$    \\
\hline
k & 0 & 2 & 2  & -1  & -1  \\
\hline
shifts $[i, j\rangle$ $(i > j)$ &  $[3, 2 \rangle$ & $[4,2 \rangle$ &  $[5,2\rangle$ &  $[6, 2 \rangle$ &  $[4, 3\rangle$    \\
\hline
k & 0 & 0   & -2 & -2 & -1 \\
\hline
shifts $[i, j\rangle$ $(i > j)$ &  $[ 5, 3\rangle$ &  $[6, 3 \rangle$ & $[ 5, 4\rangle$ &  $[6, 4\rangle$ &  $[6, 5 \rangle$  \\
\hline
k &  -2 & -2 & -2 & -2 & -1 \\
\hline\hline
\end{tabular}
\caption{The leading $z^k$  in the limit $z \to \infty$ of shifts $[i, j\rangle$ in BCFW method for $0 \to e^-(1^-)  e^+(2^+) \gamma(3^+) \gamma(4^+)  \gamma(5^-) \gamma(6^-)$ are given. }
\label{zpowerbcfw}
\end{table}
In Table (\ref{zpowerbcfw}), we tabulate the power index of $z$ in the limit $z \to \infty$ for all shifts. We call those shifts with $k < 0$ as good shifts, since the boundary terms of them vanish, and the BCFW method is expected to work.
On the other hand, for shifts with $k\ge0$, in general it is expected that they would not work.

Here we demonstrate the results of the shift $[ 1, 2  \rangle $, which has a vanishing boundary term. In the shift  $[1, 2 \rangle$, there are three independent terms which must be computed while the rest can be obtained from the boson exchanging symmetries of photons. The total amplitude can be computed and are given as below
\bea
A_{t} & = & A_P^{12} + A_Q^{12} + A_R^{12}\,,\label{nmhv1} \\
A_{P_1}^{12} & = & A_{P_3}^{12} = A_{Q_2}^{12} = A_{Q_4}^{12}=0 \,,
\\
A_{P_2}^{12} & = & \frac{S_{125}[25]\langle6|1+5|2]^2}{[15] \langle3|1+5|2] \langle3|1+2|5] \langle4|1+5|2] \langle4|1+2|5]} \label{a12p2}\,,
\\
A_{P_4}^{12} & = & \frac{S_{126}[26]\langle5|1+6|2]^2}{[16] \langle3|1+6|2] \langle3|1+2|6] \langle4|1+6|2] \langle4|1+2|6]} \,,
\\
A_{Q_1}^{12} & = & (-)\frac{ S_{123}\langle13\rangle \langle1|2+3|4]^2 }{ \langle23\rangle \langle3|1+2|5] \langle1|2+3|5] \langle3|1+2|6] \langle1|2+3|6] }  \label{a12q1} \,,
\\
A_{Q_3}^{12} & = & (-)\frac{ S_{124} \langle14\rangle \langle1|2+4|3]^2}{ \langle24\rangle \langle4|1+2|5] \langle1|2+4|5] \langle4|1+2|6] \langle1|2+4|6] } \,,
\\
A_{R_1}^{12} & = & (-)\frac{\langle15\rangle^2[24]^2 \langle1|3+5|2] }{S_{135}\langle13\rangle[26]\langle1|3+5|6] \langle3|1+5|2]} \label{a12r1}\,,
\\
A_{R_2}^{12} & = & (-)\frac{\langle15\rangle^2 [23]^2 \langle1|4+5|2] }{S_{145}\langle14\rangle [26] \langle1|4+5|6] \langle4|1+5|2] }  \,,
\\
A_{R_3}^{12} & = &(-) \frac{\langle16\rangle^2 [24]^2 \langle1|3+6|2] }{S_{136}\langle13\rangle [25] \langle1|3+6|5] \langle3|1+6|2] }  \,,
\\
A_{R_4}^{12} & = & (-)\frac{\langle16\rangle^2 [23]^2 \langle1|4+6|2] }{S_{146}\langle14\rangle [25] \langle1|4+6|5] \langle4|1+6|2] }  \,.
\eea

We have computed the amplitudes of all shifts for the process $ 0 \to e^- e^+ 4 \gamma$. It is found that the obtained amplitudes have very different forms. Since it is difficult to demonstrate the equivalence of the results given in Eq. (\ref{nmhv}) and those obtained by using BCFW shifts. Instead, we resort to numerical method to examine whether they are equivalent. We consider all allowed shifts in BCFW method, and examine whether they can yield the same results obtained from the Feynman diagram method. To do the numerical analysis, we use the Mathematica tool "S@M", which is a Mathematica Implementation of the Spinor-Helicity Formalism \cite{Maitre:2007jq}. 

\begin{table}[htb]
\centering
\begin{tabular}{|c|c|c|c|c|c|}
\hline
%Amplitudes  &  & Form Gauge independent\\
\hline
shifts $[i, j\rangle$ $(i <j)$ & $[1, 2\rangle$ &  $[1, 3\rangle$ &  $[1, 4 \rangle$ &  $[1, 5\rangle$ &  $[1, 6\rangle$    \\
\hline
&  & & & $\times \circ$  & $\times \circ$  \\
\hline
shifts $[i, j\rangle$ $(i <j)$ &  $[2, 3 \rangle$ & $[2, 4\rangle$ &  $[2, 5\rangle$ &  $[2, 6 \rangle$ &  $[3, 4\rangle$    \\
\hline
&  &   & $\times \times$ & $\times \times$  &  \\
\hline
shifts $[i, j\rangle$ $(i <j)$ &  $[3, 5\rangle$ &  $[3, 6 \rangle$ & $[4, 5\rangle$ &  $[4, 6\rangle$ &  $[5, 6 \rangle$  \\
\hline
& $\times \times$  & $\times \times$ & $\times \times$  & $\times \times$  &  \\
\hline
%Amplitudes  &  & Form Gauge independent\\
\hline
shifts $[i, j\rangle$ $(i > j)$ & $[ 2, 1\rangle$ &  $[3,1\rangle$ &  $[4, 1 \rangle$ &  $[5, 1\rangle$ &  $[6, 1\rangle$    \\
\hline
& $\surd \surd$ & $\times \times$ & $\times \times$  &  &   \\
\hline
shifts $[i, j\rangle$ $(i > j)$ &  $[3, 2 \rangle$ & $[4,2 \rangle$ &  $[5,2\rangle$ &  $[6, 2 \rangle$ &  $[4, 3\rangle$    \\
\hline
& $\times \surd$  & $\times \surd$   & & &  \\
\hline
shifts $[i, j\rangle$ $(i > j)$ &  $[ 5, 3\rangle$ &  $[6, 3 \rangle$ & $[ 5, 4\rangle$ &  $[6, 4\rangle$ &  $[6, 5 \rangle$  \\
\hline
&   &  &  &  &  \\
\hline\hline
\end{tabular}
\caption{The results of shifts $[i, j\rangle$ in BCFW method for $0 \to e^-(1^-)  e^+(2^+) \gamma(3^+) \gamma(4^+)  \gamma(5^-) \gamma(6^-)$ are given. }
\label{2f4abcfw}
\end{table}

 The numerical results are summarized in Tables \ref{2f4abcfw}, and there are a few comments.
\begin{itemize}
\item First, the shifts, like $[2, 5\rangle$, $[2, 6\rangle$, $[3, 5\rangle$, $[3, 6\rangle$, $[4, 5\rangle$, $[4, 6\rangle$, $[3,1\rangle$, and $[4, 1\rangle$, can not produce the correct results without a known boundary term. The reason lies in the fact that shifts, like $[2, 5\rangle$, $[2, 6\rangle$, $[3,1\rangle$, and $[4, 1\rangle$, change the amplitude into a form $A(z) = C_1 \frac{(z^2 + a_1 z + a_2) (z+ b) (z+c) }{(z -z_1)} + C_2 \frac{(z+b_1) (z+c_1) (z+d_1)}{(z-z_1)(z-z_2)}$, and shifts, like $[3, 5\rangle$, $[3, 6\rangle$, $[4, 5\rangle$, $[4, 6\rangle$ change the amplitude into a form $A(z) = C_1 \frac{(z+a) (z+b) }{(z-z_1)(z-z_2)} +  C_2 \frac{(z+a)(z+b)(z+c)}{(z-z_1)(z-z_2)}$, which has an either undetermined or non-vanishing boundary value in the limit $\lim_{z \to \infty} \oint \frac{A(z)}{z}$. 

\item Second, among 30 shifts of BCFW method, there are 18 shifts which satisfy the necessary condition of the BCFW method and can produce the correct results. Generally speaking, a shift with the inverse helicity of  a pair of spinors in the amplitude can always work. For example, with both fermion and photon in the amplitude, like the amplitude $A^\tree_6(1_e^-,2_{\bar{e}}^+,3_{\gamma}^-,4_{\gamma}^-,5_{\gamma}^+, 6_{\gamma}^+)$, shifts, like $[5, 4\rangle$, $[6, 4\rangle$, $[5, 3\rangle$ and $[6, 3\rangle$ can always work. Similarly, it also holds for shifts like $[1, 3\rangle$,  $[1, 4\rangle$, $[5, 2 \rangle$, and $[6, 2\rangle$.  These shifts have vanishing boundary terms.

\item Third, it is noteworthy that the shift $[2,1\rangle$ can produce the correct results, which is a little surprising. It should not be expected to work, in terms of our experience given in 4-point and 5-point amplitudes. Why this shift can work? It is found that in this shift, the amplitude given in Eq.(\ref{nmhv})  is changed into a form 
\bea
A^{21}(z) = C_1 \frac{(z^2+ a_1 z + a_2) (z + b)}{(z-z_1) (z -z_2) (z -z_3) (z -z_4)} + C_2 \frac{(z + c_1)(z + c_2)(z + c_3)}{(z - z_5) (z - z_6)(z - z_7)} \,.
\eea
In the limit $z \to \infty$, the first term proportional to $C_1$ vanishes and the second term proportional to $C_2$ apparently leads to a non-vanishing boundary value. Fortunately,  $C_2$ is found to have the following term 
\bea
\frac{1}{\langle 2 | 3+6  |1 ] } + \frac{1}{\langle 2 | 3+5 | 1 ] } + \frac{1}{\langle 2 | 4+6  | 1 ] } + \frac{1}{\langle 2 | 4+5  | 1 ] } \,, \label{2f4a12fac}
\eea
and it can be proven that this term given in Eq. (\ref{2f4a12fac}) is vanishing when it is noticed that $\langle 2 | 3+6 | 1]  = - \langle 2 | 4 + 5 | 1]$ and $\langle 2 | 3+5 | 1]  = - \langle 2 | 4 + 6 | 1]$ in term of the momentum conservation. This explains why the shift $[2, 1 \rangle$ can work.
Such kind of cancellation happens only at the full amplitude level,
which was found in Ref. \cite{Badger:2008rn}.

\item We would like to emphasize that although each of the shifts, like  $ [  1, 5  \rangle  $, $[ 1 , 6  \rangle$, $[ 3 , 2 \rangle $, and $[ 4 , 2 \rangle $, can not produce the whole results without calculating the boundary term, since under the shift the amplitude is changed into a form $C_1 + C_2 \frac{z+a}{z-z_1} +  C_3 \frac{z+ b }{(z-z_1) (z-z_2)}$. Nonetheless, the combination of two shifts, i.e. the sum of $ [ 1, 5 \rangle  $ and $[ 1 , 6  \rangle$ ( the sum of  $[ 3, 2 \rangle $  and $[ 4, 2 \rangle $) can generate the whole results. 

Below we expose more details on why the combination of  $ [  1, 5  \rangle  $ and $[ 1 , 6  \rangle$ can produce the full results. Under the shift $[1, 5 \rangle$, the amplitude given in Eq. (\ref{nmhv}) can be put into a form 
\bea
A^{15}(z) &=& C_1^{15} +  C_2^{15} \frac{z + \frac{\spb1.2}{\spb5.2} }{ z + \frac{\spb1.6}{\spb5.6} } +  \left ( C_3^{15} \frac{z - \frac{\langle 5 | 2 + 3 | 4] }{ \langle 1 | 2 + 3 | 4] }}{ (z + \frac{\spb1.6}{\spb5.6}) (z - \frac{S_{235}}{\langle 1 | 2+ 3| 5] })}  + (3 \leftrightarrow 4) \right ) \,,
\eea
\bea
C_1^{15} &=& \frac{1}{ \spa1.4  \spb2.6} \frac{ \spa1.5 \spb2.3}{\spb1.5 \spa2.3 } \frac{\langle 6 |2+3 | 4 ] }{S_{236}}  +   \frac{1}{ \spa1.3  \spb2.6} \frac{ \spa1.5 \spb2.4}{\spb1.5 \spa2.4 } \frac{ \langle 6 |2+4 | 3 ] }{S_{246}} \,,
\eea
\bea
C_2^{15} &=& \frac{\spb5.2 \langle 1 | 5+6 |2] S_{156} }{  \spa1.3 \spa2.3 \spa1.4 \spa2.4 \spb1.5 \spb2.5 \spb2.6 \spb5.6 } \,,
\eea
\bea
C_3^{15} &=& \frac{\langle 1 | 2 + 3 | 4] \spa1.6 \spa2.3 }{ \spb5.6 \spa1.4 \spa2.3 \spb2.5 \langle 1 | 2+ 3 | 5] } \,.
\eea
The residue of $z^{15} \to \infty$ can be found 
\bea
 Res_{z^{15} \to \infty} (\frac{A^{15}(z)}{z}) = -  C_1^{15}  - C_2^{15} = - B^5\,, \label{2f4a15}
\eea
where $B^5$ denotes the boundary term. Under the shift $[1, 6\rangle$, the amplitude is changed into the following form
\bea
A^{16}(z) &=& C_1^{16} +  C_2^{16} \frac{z + \frac{\spb1.2}{\spb6.2} }{ z - \frac{\spb1.5}{\spb5.6} } +  \left ( C_3^{16} \frac{z - \frac{\langle 6 | 2 + 3 | 4] }{ \langle 1 | 2 + 3 | 4] }}{ (z - \frac{\spb1.5}{\spb5.6}) (z - \frac{S_{236}}{\langle 1 | 2+ 3| 6] })}  + (3 \leftrightarrow 4) \right ) \,,
\eea
\bea
C_1^{16} &=& \frac{1}{ \spa1.4  \spb2.5} \frac{ \spa1.6 \spb2.3}{\spb1.6 \spa2.3 } \frac{ \langle 5 |2+3| 4 ]  }{S_{235}} + \frac{1}{ \spa1.3  \spb2.5} \frac{ \spa1.6 \spb2.4}{\spb1.6 \spa2.4 } \frac{ \langle 5 |2+4 | 3 ] }{S_{245}}\,,
\eea
\bea
C_2^{16} &=& - \frac{\spb6.2 \langle 1 | 5+6 |2] S_{156} }{  \spa1.3 \spa2.3 \spa1.4 \spa2.4 \spb2.5 \spb1.6 \spb2.6 \spb5.6 } \,,
\eea
\bea
C_3^{16} &=& - \frac{\langle 1 | 2 + 3 | 4] \spa1.5 \spa2.3 }{ \spb5.6 \spa1.4 \spa2.3 \spb2.6 \langle 1 | 2+ 3 | 6] } \,.
\eea
The residue of $z^{16} \to \infty$ can be found 
\bea
Res_{z^{16} \to \infty} (\frac{A^{16}(z)}{z}) = -  C_1^{16}  - C_2^{16} = - B^6\,, \label{2f4a16}
\eea
where $B^6$ denotes the boundary term.

From Eq. (\ref{2f4a15}) and Eq. (\ref{2f4a16}), it is observed that 
\bea
Res_{z^{15} \to \infty} (\frac{A^{15}(z)}{z})   + Res_{z^{16} \to \infty} (\frac{A^{16}(z)}{z})   = - A^{tree}_6(1^-_e, 2^+_{\bar{e}}, 3^+_\gamma,4^+_\gamma,5^-_\gamma,6^-_\gamma)\,.
\eea
Therefore, by using the analyticity of $A^{15}(z)/z$ and $A^{16}(z)/z$ and the fact $A^{15}(0) = A^{16}(0) = A^{full}$, we arrive at the result
\bea
A^{15}(0 ) + A^{16}(0) -B^5 - B^6 = A^{full} =  - Res_i(\frac{A^{15}(z)}{z}) - Res_i(\frac{A^{16}(z)}{z}) \label{mAsum} \,, 
\eea
which explains why the sum of the shifts $[1, 5\rangle$ and $[1, 6\rangle$ produces the whole result. Similar reasoning also holds for the sum of shifts $[3, 2 \rangle$ and $[4, 2\rangle$.

Since the sum of the shifts  $[1 , 5 \rangle  $ and  $[1 , 6 \rangle  $ can produce the full amplitude result, therefore we can put the total amplitude from the amplitudes obtained from these two shifts as
\bea
A_{t} = [P(5,6)] \left ( \frac{S_{156}^2\langle2|1+6|5]}{\langle23\rangle \langle24\rangle [16][56] \langle3|1+6|5]  \langle4|1+6|5]} - [P(3,4)] \frac{\langle16\rangle^2[24]^2}{S_{136}\langle13\rangle [25] \langle3|1+6|5] }\ \right )
\eea
In term of the number of term for the full amplitude, it reads as $2 \times (1 + 2) = 6$. Although the form of this amplitude is different from the one given in Eq. (\ref{nmhv}), we have tested that numerically that they are equal. Meanwhile, this form of amplitude has an explicit property that it is the sum of two terms which is unchanged under the shifts $[1, 5 \rangle$ and $[1, 6\rangle$, respectively.

\end{itemize}

It is found that although each of the shifts $[1, 5\rangle$, $[1, 6\rangle$, $[3, 2\rangle$, and $[4, 2\rangle$, could not yield the full amplitude without evaluating the boundary term, the sum of the shifts $[1, 5\rangle$ and $[1, 6\rangle$ ($[3, 2\rangle$ and $[4, 2\rangle$) indeed can produce the whole amplitude. Besides, the calculation procedure for these shifts in the BCFW method is quite simple. So it is worthy to have a close look at the calculation procedure of them. For example, for the shift $[1 , 5 \rangle  $, the total amplitude is given as below
\bea
A_{t} & = & A_{P}^{15} + A_{Q}^{15} + A_{R}^{15}\,,\label{nmhv7}\\
A_{P_1}^{15} & = & A_{P_2}^{15}=A_{P_3}^{15} =A_{Q}^{15}= A_{R_1}^{15}= A_{R_2}^{15}=0\,,\\
A_{P_4}^{15} & =&\frac{S_{156}^2\langle2|1+6|5]}{\langle23\rangle \langle24\rangle [16][56] \langle3|1+6|5]  \langle4|1+6|5]} \label{a15p1}\,,\\
A_{R_3}^{15} &=& (-)\frac{\langle16\rangle^2[24]^2}{S_{136}\langle13\rangle [25] \langle3|1+6|5] }\,,\\
A_{R_4}^{15} &=& (-)\frac{\langle16\rangle^2[23]^2}{S_{146}\langle14\rangle [25] \langle4|1+6|5] }\,.
\eea
There are only two independent terms needed to be computed. For example, according to the diagram given in Figure 7, the amplitude $A_{P_4}^{15}$ can be put as
\bea
A_{P_4}^{15}  &= & \frac{ \spa1.6^2 }{\spa1.{\hat{p}}} \frac{1}{ S_{16} } \frac{ \spa{-\hat{p}}.{\hat{5}}^2 \spa{-\hat{p}}.2}{\spa{-\hat{p}}.3 \spa{-\hat{p}}.4 \spa2.3 \spa2.4} \,, \label{a151}
\eea
from the pole condition it is easy to find that the shifted spinors, which can be solved as 
\bea
| \hat{5} \rangle &=&  \frac{ ( 5 + 1 ) | 6 ] }{\spb6.5}\,,\\ 
| \hat{p} \rangle &=& \frac{[ (1 + 6)|5] }{\spb5.6}\,. 
\eea
Substituting these two spinors into Eq. (\ref{a151}), we arrive at Eq. (\ref{a15p1}). Moreover, it is observed that there are 5 brackets with shifted momenta needed to be computed.

From the diagram given in Figure 8, the amplitude $A_{R_3}^{15}$ can be put as 
\bea
A_{R_3}^{15} & = & \frac{ \spa1.6^2  }{ \spa1.4 \spa{\hat{r}}.4 } \frac{1}{S_{146}} \frac{\spb2.3^2}{\spb{-\hat{r}}.5 \spb2.5 }\,, \label{a152}
\eea
from the pole condition it is easy to find that the shifted momentum $\hat{r}$, which is given as 
\bea
\hat{r} &= & |1 \rangle [1 | + |4 \rangle [4 | + |6 \rangle [6 | -\frac{S_{146} }{ \langle 1 | 4 + 6 | 5 ] }  |1 \rangle [5 | \,,
\eea
and there is no complicate calculation to evaluate $\langle 4 | \hat{p} | 5]$ and the calculation is straightforward 
\bea
\langle 4 | \hat{r} | 5] & = & \langle 4 | 1 + 6 | 5]\,.
\eea
Meanwhile, there are only 2 brackets with shifted momenta needed to be computed. While amplitude $A_{R_4}^{15}$ can be obtained by using the boson exchanging symmetry between $3 \leftrightarrow 4$, which should be quite simple. Another interesting fact is that the amplitude $A_t^{15}$ is unchanged if we make a shift $[1, 6\rangle$ on it. 

Similarly, under the shift $[1 , 6 \rangle  $, the total amplitude can be directly computed from diagrams and is given as below
\bea
A_{t} & = & A_{P}^{16} + A_{Q}^{16} + A_{R}^{16}\,,\label{nmhv9}\\
A_{P_1}^{16} & = & A_{P_3}^{16}=A_{P_4}^{16} =A_{Q}^{16}= A_{R_3}^{16}= A_{R_4}^{16}=0\,,\\
A_{P_2}^{16} & =&(-)\frac{S_{156}^2\langle2|1+5|6]}{\langle23\rangle \langle24\rangle [15][56] \langle3|1+5|6]  \langle4|1+5|6]}\,,\\
A_{R_1}^{16} &=& (-)\frac{\langle15\rangle^2[24]^2}{S_{135}\langle13\rangle [26] \langle3|1+5|6] }\,,\\
A_{R_2}^{16} &=& (-)\frac{\langle15\rangle^2[23]^2}{S_{145}\langle14\rangle [26] \langle4|1+5|6] }\,.
\eea
It is observed that the results of $[1 , 6 \rangle  $ shift can be directly obtained from the results of $[1, 5\rangle$ shift by using the exchanging symmetry $5 \leftrightarrow 6$. Meanwhile, obviously, the amplitude $A_t^{16}$ is unchanged if we perform a $[1, 5\rangle$ shift on it.

 \begin{table}[htb]
 \begin{center}
 \begin{tabular}{|c|c|c|c|}
\hline
& I.T. & No. of Calculation for I.T. & No. of terms for F.A.  \\
\hline
$[1,2\rangle$ & 3 & 16 &  8 \\
\hline
$[1,3\rangle$ & 3 & 15 &  5\\
\hline
$[5,6\rangle$ & 4 & 16 & 6 \\
\hline
$[1,5\rangle$ + $[1,6\rangle$   &  2 & 7 &  6\\
\hline
\end{tabular}
\caption{The comparison on the independent terms (I.T), number of calculation of shifted brackets in all independent terms, and the number of terms for the full amplitudes (F.A.),  for shifts $[1, 2\rangle$, $[1,3\rangle$, $[5, 6 \rangle$, and the sum of shifts $[1,5\rangle$ and $[1,6\rangle$ are provided. }
\label{sixm}
\end{center}
 \end{table}

It is interesting to compare the computation procedure of the sum of $[1, 5\rangle$ and $[1, 6\rangle$ shifts with some good shifts like the $[1, 2 \rangle$ shift, $[1 ,3 \rangle$ shift, and $[5,6\rangle$ shift, as presented in Table (\ref{sixm}). There are a few comments on this comparison below. 
\begin{itemize}
\item 1) In the $[1 , 2 \rangle$ and $[1, 3\rangle$, there are three independent terms which should be computed before using the boson exchanging symmetries. In the shift $[5,6\rangle$, there are 4 independent terms. While for the sum of $[1, 5\rangle$ and $[1, 6\rangle$ shifts, there are only 2 independent terms which should be computed. 
\item 2) Meanwhile, the total number of terms of the amplitude in $[1, 2 \rangle$ ($[1, 3 \rangle$) shift is 8(5). The total number of terms is 6 for the shift $[5, 6\rangle$. While the total number of terms in the $[1, 5\rangle$ and $[1, 6\rangle$ shifts is 6. 
\item 3) The third,  as shown above, there are more terms, like spinors $|\hat{p}\rangle$, $|\hat{q}]$, and $\hat{r}$, which need to be computed for $[1, 2 \rangle$, $[1, 3 \rangle$, and $[5, 6\rangle$ shifts than for the $[1, 5\rangle$ and $[1, 6\rangle$ shifts. This may save the CPU time in a realistic computation.
\end{itemize}
Therefore, from this comparison, we can conclude that the $[1, 5\rangle$ and $[1, 6\rangle$ shifts could be more economic than the $[1, 2 \rangle$ shift, $[1, 3 \rangle$ shift, and $[5, 6\rangle$. Similarly, we also find that the sum of $[3, 2\rangle$ and $[4, 2\rangle$ shifts can also produce the result of the full amplitude.

To separate such a shift from other shifts which have vanishing boundary terms, we call this shift as the LLYZ shift.
 
 \section{Proof of LLYZ shift for more general processes}

We have examined this novel shift can work for the helicity amplitudes of the process $0 \to e^- e^+ 5 \gamma $ and $0 \to e^- e^+ 6 \gamma $ where the amplitudes can be worked out explicitly as given in section III from Eq.(\ref{nmhv7t}) to Eq. (\ref{a8nn6}). For example, for the NMHVA of the process $0 \to e^- e^+ 5 \gamma $ and $0 \to e^- e^+ 6 \gamma$, there are two photons with negative helicity, we have 
\bea
\sum_{\gamma_i^-} B^i= \sum_{\gamma_i^-} - Res_{z \to \infty} \frac{A^{[1, \gamma^-_i\rangle } }{z} =  A^{full}\,,
\eea
Then the sum of these two shifts indeed can produce one full amplitude. 

For more general case, we can denote the amplitudes of the processes  $0\to e^- e^+ \gamma_1^+ \cdots \gamma_n^+ \gamma_1^- \cdots \gamma_m^-$ with $ 1 < m \leq n$ as $A_{n,m}$, where $n$ and $m$ represent the number of photons with positive and negative helcities respectively. We can conjecture that in this novel shift that the sum of the shifts $[1, \gamma_i^-\rangle$ in the limit $z \to \infty$ produces a negative full amplitude i.e. we have
\bea
\sum_{\gamma_i^-} B^i= \sum_{\gamma_i^-} - Res_{z \to \infty} \frac{A^{[1, \gamma_i^-\rangle} (z)}{z} = A_{n,m}^{full} \,.
\eea
Thus, the sum of these shifts $[1^-, \gamma_i^-\rangle$ can produce a result $(m-1) \,A_{n,m}^{full}$, where each amplitudes of each shift can be obtained by using the BCFW method. Such a result can be tested explicitly by using the actual $N^2$MHV amplitude given in Eq. (\ref{nmhv8n2}).

For the general amplitudes of the processes  $0\to e^-(1^-) e^+(2^+) \gamma_3^{h_3} \cdots \gamma_n^{h_n}$, we can prove that such an identity can hold by choosing the BG gauge. In terms of spinor convention \cite{Badger:2008rn}, the most general QED tree-level amplitudes \cite{Kleiss:1986qc} obtained by using Feynman diagrams can be put as
\begin{align}
    A_{full}=&\frac{1}{\prod_{j=3}^{n}\za{p_{ref}^{j,h_j} | j^{h_j}}}\sum_{\sigma\in S_{n-2}}F(1,2;\sigma(3)^{h_{\sigma_3}},\cdots,\sigma(n)^{h_{\sigma_n}}) \,,\\
    F(1,2; 3^{h_3},\cdots,n^{h_n})=&\za{a_3 \,1}\zb{2 \, b_n}\prod_{i=3}^{n-1}\frac{\bra{a_{i+1}} 1+K_{3,i}|{b_{i}}]}{(1+K_{3,i})^2} \,,
\end{align}
where $K_{3,i}=\sum_{j=3}^{i}k_j$, and $\za{p_{ref}^{j,h_j} | j^{h_j}}$ is originated from the wave functions of photons and can be understood as $\langle p^- | q^+ \rangle = \spa{p}.{q}$ and $\langle p^+ | q^- \rangle = \spb{p}.{q}$ where $p_{ref}^{j,h_j}$ is called as reference momentum, which is an arbitrary light-like four vector and is dependent upon the momentum and helicity of photon numbered as $j$. Obviously, in such a form, the boson exchanging symmetries are explicit as built in the sum of all permutations.

Generally, in the BG gauge, the reference momentum $p_{ref}^{j, h_j}$ for the photon numbered as $j$ can be put as
\bea
p_{ref}^{j, h_j} = \frac{1 + h_j}{2} p^1+ \frac{1 - h_j}{2} p^2\,,
\eea
where $p^1$ is the momentum of $e^-$ (labelled as $1$ for the sake of simplicity when no confusion can be arisen) and $p^2$ is the momentum of $e^+$ (labelled as $2$ for the sake of simplicity). Then for all photons with a positive helicity, we have $p_{ref}^{j,-}=p_1$ and for all photons with a negative helicity, we have $p_{ref}^{j,+}=p_2$. Therefore, under such a convention, the momenta of spinors $a_i$ and $b_i$ (with $3 \leq i \leq n$) in $F$ can be defined 
defined as
\bea
a_i &=& \frac{1 + h_i}{2} p_{ref}^{i, h_i}+ \frac{1 - h_i}{2} k_i\,, \\
b_i &=& \frac{1 + h_j}{2} k_i+ \frac{1 - h_j}{2} p_{ref}^{i,h_i} \,.
\eea
Then we can arrive at that if $h_i=+1$ then $a_i=p^1$ and $b_i=k_i$ as well as if $h_i=-1$ then $a_i=k_i$ and $b_i=p^2$. Below, for the sake of simplicity, we will use $1$ ($2$) to denote $p^1$ ($p^2$).

With this convention, we have:
\begin{align}
    F(1^-, 2^+; i^{+},\cdots)=0\,,
\end{align}
simply because of $\spa{a_i}.1=\spa1.1=0$.

Considering the contribution of a particular helicity configuration, without loss of generality, 
we label it as $F(1^-,2^+;3^-,4^-,\cdots,(j-1)^-,j^+,(j+1)^{h_{j+1}},\cdots,n^{h_n})$,
where $j$ is the first $\gamma$ that has positive helicity,
i.e. $h_3=h_4=\cdots=h_{j-1}=-$.
Summing over all permutation of particles from $3^-$ to $(j-1)^-$,
%, which can be labelled as $A_{hel}$. Obviously, we can have
%\bea
%A^{full} = \sum_{hel} A_{hel}\,,
%\eea
%since the sum goes over all helicity configurations and in a given helicity configuration all permutations of photons with the same helicity, either positive or negative. It is helpful to refer the helicity amplitudes given from Eq. (\ref{nmhv7t}) to Eq.(\ref{a8nn6}) to understand this.Then by definitions, the helicity amplitude $A_{hel}$ 
we obtain the following results:
\begin{align}
    F((j+1)^{h_{j+1}},\cdots,n^{h_n})=&\sum_{\sigma^-}F(1^-,2^+;\sigma(3)^-,\cdots,\sigma(j-1)^-,j^+,R)\\
    =&\frac{[12]^{j-4}}{\prod_{i=3}^{j-1} [1i]}\bra{1} \sum_{i=3}^{j-1}k_{i}| 2] \spb{b_n}.2  \prod_{i=j}^{n}\frac{\bra{a_{i+1}} 1 +K_{3,i}|b_i]}{( 1+K_{3,i})^2} \,.
\end{align}
Therefore, in $[1,3^-\rangle$ shift, such contribution to the amplitude is modified as:
\begin{align}
    A^{[1,3\rangle}((j+1)^{h_{j+1}},\cdots,n^{h_{n}}) = \frac{1}{\prod_{i=3}^{n} \za{p_{ref}^{i,h_i} | i^{h_i} } } \frac{[\hat{1}2]^{j-4}}{\prod_{i=3}^{j-1} [\hat{1}i]}\bra{\hat{1}}  \sum_{i=3}^{j-1}\hat{k}_{i}| 2] \spb{b_n}.2 \prod_{i=j}^{n}\frac{\bra{\hat{a}_{i+1}} \hat{1}+\hat{K}_{3,i}|b_i]}{(\hat{1}+\hat{K}_{3,i})^2}\,,
\end{align}
where $\hat{a}_{i+1}$ can be $\hat{1}$ when the helicity of $h_i=-1$. At the limit $z\to\infty$, the amplitude can be put in the following form
\begin{align}
    \lim_{z \to \infty} A^{[1,3\rangle}((j+1)^{h_{j+1}},\cdots,n^{h_n}) = & \frac{1}{\prod_{i=3}^{n} \za{p_{ref}^{i,h_i} | i^{h_i} } } \frac{\spb1.2^{j-4}}{\prod_{i=3}^{j-1}\zb{1n}} T^{[1,3\rangle} \bra{1} \sum_{i=3}^{j-1}k_{i}| 2]  \spb{b_n}.2 \prod_{i=j}^{n}\frac{\bra{a_{i+1}} 1+K_{3,i}|b_i]}{(1+K_{3,i})^2} \nonumber \\
                                                                        =& A((j+1)^{h_{j+1}},\cdots,n^{h_n}) \,\,T^{[1,3\rangle}\,, 
\end{align}
where we have used the fact $\hat{1} +\hat{K_{3,i}} = 1 +K_{3,i}$ and $\bra{\hat{1}} \sum_{i=3}^{j-1} \hat{k}_i | 2] = \bra{1} \hat{1}+\sum_{i=3}^{j-1} \hat{k}_i | 2] = \bra{1} \sum_{i=3}^{j-1} k_i | 2]$. Meanwhile the factor $T^{[1, 3\rangle}$ can be found as 
\bea
T^{[1,3\rangle}&=& \frac{\prod_{i=3}^{j-1} \spb1.i }{[\spb1.2^{j-4}} \frac{[32]^{j-4}}{[13]\prod_{i=4}^{j-1}[3i]} \,.
   \eea
Summing over all such shifts, we can get:
\bea
\sum_{m=3}^{j-1} \lim_{z \to \infty} A^{[1,m\rangle}((j+1)^{h_{j+1}},\cdots,n^{h_n}) = A((j+1)^{h_{j+1}},\cdots,n^{h_n}) \sum_{m=3}^{j-1} T^{[1, \gamma_m^-\rangle}\,.
\eea
Using the Schouten identity and mathematical induction as given in Eq. (\ref{bcfwnp}), we can obtain
\bea
    \sum_{m=3}^{j-1} T^{[1, \gamma_m^-\rangle} &=& \frac{\prod_{i=3}^{j-1} \spb1.i }{\spb1.2^{j-4}}  (\sum_{m=3}^{j-1}\frac{[m2]^{j-4}}{[1m]\prod_{i=3,i\ne m}^{j-1}[mi]}) \nonumber \\
    &=& \frac{\prod_{i=3}^{j-1} \spb1.i }{\spb1.2^{j-4}} \frac{[12]^{j-4}}{\prod_{i=3}^{j-1}[1i]}  \nonumber \\
    &=& 1\,.
\eea
Furthermore, we notice that for a shift $[1,\gamma_i^-\rangle$ with $i>j$, it vanishes in the limit at $z\to \infty$.
So we arrive at the result
\begin{align}
    \sum_{\gamma_j^{-}} \lim_{z\to\infty}A^{[1,\gamma_j^-\rangle}((j+1)^{h_{j+1}},\cdots,n^{h_n}) =A((j+1)^{h_{j+1}},\cdots,n^{h_n})\,.
\end{align}
Summing over all possible choices, we conclude that
\begin{align}
    \sum_{\gamma_j^{-}} \lim_{z\to\infty}A^{[1,\gamma_j^-\rangle}_{full}=A_{full} \label{summed}\,.
\end{align}

According to our proof given above, the identity given in Eq. (\ref{summed}) is tightly related to the property of helicity amplitudes in the BG gauge, i.e. that only diagrams with the fermion $1^-$ adjoint with a photon with negative helicity can contribute to the total amplitude. While total amplitude can always be put as the sum of terms $B^i$ ,which denotes the contribution of all diagrams with $f (1^-)$ adjoint with the photon $\gamma_i^-$. Similarly,  the total amplitude can also be put as the sum of terms $\bar{B}^i$, which denotes the contributions of diagrams where  $\bar{f} (2^+)$ adjoint with a photon with a positive helicity $\gamma_i^+$.

\section{Comparison of the LLYZ shift with other shifts in the BCFW method}

In order to avoid the quick increase of number of terms in the amplitude, in the LLYZ shift, we always choose the sum of $[1^-, \gamma_i^-\rangle$ shifts, instead of the sum of $[\gamma_i^+, 2 \rangle$, which should also work for  the processes $0\to e^- e^+ \gamma_1^+ \cdots \gamma_n^+ \gamma_1^- \cdots \gamma_m^-$ with $ m \leq n$. By using the BCFW method, the number of terms in the NMHV amplitudes of the process $0\to e^- e^+ \gamma_1^+ \cdots \gamma_n^+ \gamma_1^- \gamma_2^-$  in the LLYZ shift can be found as
\bea
T_{N}(n) = 2 (2^n -1)\,,
\eea
where the factor $2$ represents the fact that there are two shifts for the sum, i.e. $[1, \gamma_1^- \rangle$ and $[1, \gamma_2^- \rangle$. Similarly, the number of terms in the N$^2$MHV amplitudes of the process $0\to e^- e^+ \gamma_1^+ \cdots \gamma_n^+ \gamma_1^- \gamma_2^- \gamma_3^-$  in the LLYZ shift can be given as
\bea
T_{N^2}(n) = 3 \times  (6\times3^n - 8 \times 2^n - 3 \times n + 2) \,,
\eea
where the factor $3$ reflects the fact that there are threee shifts for the sum, i.e. $[1, \gamma_1^- \rangle$, $[1, \gamma_2^- \rangle$, and $[1, \gamma_3^- \rangle$.The number of terms in the N$^3$MHV amplitudes  $0\to e^- e^+ \gamma_1^+ \cdots \gamma_n^+ \gamma_1^- \gamma_2^- \gamma_3^-  \gamma_4^-$  in the LLYZ shift can be given as
\bea
T_{N^3}(n) &=& 4 \times  \left[  3 \times \left ( \sum_{k=0}^{n-3} C_n^k \,\, T_{N^2}(n-k) + C_n^{n-2}  \,\,T_{N}(3)  + C_n^{n-1} \right ) \right. \nonumber \\
& & + \left. 3 \times \left ( 2 C_{n}^1 \,\,T_{N}(n-1) +  \sum_{k=2}^{n-2} C_{n}^k  \,\,T_{N}(k) \,\,T_{N}(n-k) \right ) \right. \nonumber \\
& & + \left. \left (  C_n^{1}+ C_n^{2} \,\,T_{N}(3) +  \sum_{k=3}^{n-1} C_{n}^{k} \,\, T_{N^2}(k) \right ) \, \right ], \label{llyz:TN3}
\eea
where the overall factor $4$ indicates there are four shifts to be summed. While each line represents one topology and $C_n^k=\frac{n!}{k! (n-k)!}$. In principle, for other higher MHV amplitudes like N$^4$MHV, N$^5$MHV and so on, based on the BCFW method, we can derive the recursion relations of the number of terms.

It is also interesting to compare the number of terms of different shifts in the BCFW method for the NMHV amplitudes for the process $0 \to e^-(1^-) e^+(2^+)  \gamma_1^+ \cdots \gamma_n^+ \gamma_1^- \gamma_2^-$, as given in Table \ref{notinbcfwnmhva}.
\begin{table}[htb]
\centering
\begin{tabular}{|c|c|c|c|c|c|c|c|c|c|}
\hline
%Amplitudes  &  & Form Gauge independent\\
\hline
$N_p$  & 6 & 7 & 8 & 9 & 10 & 11 & 12 & 13 & 14\\
\hline
$N_\gamma$  & 4 & 5 & 6 & 7 & 8 & 9 & 10 & 11 & 12 \\
\hline
$N_{\gamma^+}$  & 2 & 3 & 4 & 5 & 6 & 7 & 8 & 9 & 10 \\
\hline
\hline
LLYZ shift  &  6 & 14 & 30 & 62 & 126 &  254 & 510 & 1022 & 2046 \\
\hline
BCFW(Dressed)& 6 &14 &30&62 & 126 & 254 & 510 & 1022 & 2046 \\
\hline
$[\gamma_1^-,\gamma_2^-\rangle$ &  6 & 14 & 30 & 62 & 126 &  254 & 510 & 1022 & 2046 \\
\hline
$[1^-, \gamma^+ \rangle $ & 5 & 13 & 29 & 61 & 125 & 253  & 509& 1021 & 2045 \\
\hline
\hline
$[\gamma_i^+,\gamma_j^+\rangle $  & 6 & 20 & 56 & 144 & 352 & 832  & 1920 & 4352  & 9728 \\
\hline
$[\gamma_1^-,\gamma_i^+\rangle $  & 6 & 20 & 56 & 144 &  352 & 832  & 1920 & 4352 & 9728\\
\hline
$[\gamma^-,2^+\rangle $ & 5 & 22 & 103 & 546 & 3339 & 23500  & 188255 & 1694806 & 16949083 \\
\hline
$[1^-,2^+\rangle$ & 8 & 38 & 182 & 972 & 5958 & 41960  & 336190 & 3026732 &  30269366\\
\hline
\hline
\end{tabular}
\caption{Number of terms of NMHV amplitudes for $0\to e^- e^+ \gamma_1^+ \cdots \gamma_n^+ \gamma_1^- \gamma_2^- $ with a few typical shifts in the BCFW method are provided. The dressed BCFW method is proposed in \cite{Badger:2010eq}. }
\label{notinbcfwnmhva}
\end{table}
It is observed that the number of terms in the amplitude of the LLYZ shift is the same as the $[\gamma_1^-, \gamma_2^-\rangle$. The number of terms for $[1^-, \gamma^+\rangle$ is smaller than LLYZ shift by one for all $N_{\gamma^+}$. 

We also provide a comparison of the number of terms of a few typical shifts in the BCFW method for the process $0 \to e^- e^+  \gamma_1^+ \cdots \gamma_n^+ \gamma_1^- \gamma_2^- \gamma_3^-$, as given in Table \ref{notinbcfwn2mhva}.
\begin{table}[htb]
\centering
\begin{tabular}{|c|c|c|c|c|c|c|c|}
\hline
%Amplitudes  &  & Form Gauge independent\\
\hline
$N_p$ & 8 & 9 & 10 & 11 & 12 & 13 & 14\\
\hline
$N_\gamma$  & 6 & 7 & 8 & 9 & 10 & 11 & 12 \\
\hline
$N_{\gamma^+}$  & 3 & 4 & 5 & 6 & 7 & 8 & 9 \\
\hline
LLYZ shift   & 273 & 1044 & 3567 & 11538  & 36237 & 111888 & 341931 \\
\hline
$[\gamma_1^-,\gamma_2^-\rangle$  & 112 & 444 & 1544 &  5044 & 15936 & 49388 & 151288 \\
\hline
$[1^-,\gamma^+\rangle$  & 106 & 451 & 1624 &  5425 & 17398 & 54463 & 167956 \\
\hline
\hline
$[\gamma_i^+,\gamma_j^+\rangle $  & 112  & 620 & 2860 & 11876  & 46108  & 170948 & 613084 \\
\hline
$[\gamma^-,\gamma^+\rangle $  & 138 & 710 & 3150  & 12782  & 48894 & 179438 & 638814\\
\hline
$[\gamma^-,2^+\rangle$  & 142 & 1037 & 8101 & 67971  & 617275 & 6088429 & 65204413  \\
\hline
$[1^-,2^+\rangle$  & 390 & 3330 & 27816 & 241458  & 2236044 & 22329138 & 241156248 \\
\hline
\end{tabular}
\caption{Number of terms of N$^2$MHV amplitudes for $0\to e^- e^+  \gamma_1^+ \cdots \gamma_n^+ \gamma_1^- \gamma_2^- \gamma_3^-$ with a few typical shifts in the BCFW method are provided.}
\label{notinbcfwn2mhva}
\end{table}
It is found that the shift $[1^-, \gamma^+\rangle$ has the least number of terms in the full amplitude for the N$^2$MHV amplitude for the process $0\to e^- e^+  \gamma_1^+ \cdots \gamma_n^+ \gamma_1^- \gamma_2^- \gamma_3^-$ when $N_{\gamma^+}=3, 4$, the shift $[\gamma^-, \gamma^- \rangle$ has the least number of terms when $N_{\gamma^+} > 5$.

Since the shift with a pair of photons with negative helcities has least number of terms, we also provide the relevant formula to count it. The number of the terms in the amplitude for the shift $[\gamma_1^-, \gamma_2^-\rangle$ for the process $0\to e^- e^+  \gamma_1^+ \cdots \gamma_n^+ \gamma_1^- \gamma_2^- \gamma_3^- $ can be put as
\bea
T^{\gamma^-\gamma^-}_{N^2}(n) = 4 \left ( \sum_{i=2}^{n-1} \,C_n^i \,T_N(i) + C_n^1 \right) + 2 T_N(n)\,.
\eea
The number of the terms in the amplitude for the shift $[\gamma_1^-, \gamma_2^-\rangle$ for the process $0\to e^- e^+  \gamma_1^+ \cdots \gamma_n^+ \gamma_1^- \gamma_2^- \gamma_3^- \gamma_4^-$ can be put as
\bea
T^{\gamma^-\gamma^-}_{N^3}(n) &=& 4 \left ( \sum_{i=1}^{n-3} \,C_n^i \,T_{N^2}^{\gamma^-\gamma^-}(n-i) + C_n^2  \, T_N(3) + C_n^1 \right) + 2 T_{N^2}^{\gamma^-\gamma^-}(n)\,, \nonumber \\
 & & + 4 \left ( \sum_{i=2}^{n-2} \,C_n^i \,T_{N}(i) \, T_{N}(n-i) + 2 C_n^1 \, T_N(n-1) \right)\,.
\eea

\begin{table}[htb]
\centering
\begin{tabular}{|c|c|c|c|c|c|}
\hline
%Amplitudes  &  & Form Gauge independent\\
\hline
$N_p$ & 6 & 8 &  10 &  12 &  14 \\
\hline
$N_\gamma$ & 4  & 6 & 8 &  10 & 12 \\
\hline
$N_{\gamma^+} = N_{\gamma^-} $ & 2 & 3 & 4 & 5 & 6 \\
\hline
LLYZ shift   & 6  & 273  &  35344 & 9793805  & 4921520256 \\
\hline
$[\gamma_1^-,\gamma_2^-\rangle$  & 6 & 112 & 4344 & 144178  & 23608048 \\
\hline
$[1^-,\gamma^+\rangle$  & 5  & 106  & 5041 & 424592 &  55741427 \\
\hline
\end{tabular}
\caption{Number of terms of N$^{\frac{N}{2}}$MHV amplitudes for the process $0\to e^- e^+  \gamma_1^+ \cdots \gamma_n^+ \gamma_1^- \cdots \gamma_n^-$ with a few typical shifts in the BCFW method are provided.}
\label{nn2bcfw}
\end{table}

For the shift  $[1^-,\gamma^+\rangle $, the number of terms in the NMHV amplitudes of the process $0\to e^- e^+ \gamma_1^+ \cdots \gamma_n^+ \gamma_1^- \gamma_2^-$  can be expressed as
\bea
T_{N}^{1\gamma^-}(n) = 2^{n+1} -3\,. 
\eea
Similarly, the number of terms in the N$^2$MHV amplitudes of the process $0\to e^- e^+ \gamma_1^+ \cdots \gamma_n^+ \gamma_1^- \gamma_2^- \gamma_3^-$  in the $[1^-,\gamma^+\rangle $ shift can be given as
\bea
T_{N^2}^{1\gamma^-}(n) = T_{N^2}^{1\gamma^-}(n-1)+3\left (n+\sum_{i=0}^{n-3}C_{n-1}^i T_{N}^{1\gamma^-}(n-1+i)\right)+3\left (1+\sum_{i=1}^{n-1}C_{n-1}^iT^{1\gamma^-}_{N}(i+1)\right)\,.
\eea
The number of terms in the N$^3$MHV amplitudes  $0\to e^- e^+ \gamma_1^+ \cdots \gamma_n^+ \gamma_1^- \gamma_2^- \gamma_3^-  \gamma_4^-$  in the $[1^-,\gamma^+\rangle $ shift can be given as
\bea
T^{1\gamma^-}_{N^3}(n) &=& T^{1\gamma^-}_{N^3}(n-1)+4\left (n-1+C_{n-1}^{2}T^{1\gamma^-}_{N}(3)+\sum_{i=0}^{n-3}C_{n-1}^i T^{1\gamma^-}_{N^2}(n-1-i)\right) \nonumber\\
& & + 6\left (nT^{1\gamma^-}_{N}(n-1)+T^{1\gamma^-}_{N}(n)+\sum_{i=1}^{n-3}C_{n-1}^i T^{1\gamma^-}_{N}(i+1)T^{1\gamma^-}_{N}(n-1-i)\right) \nonumber\\
& & + 4 \left (  1+(n-1)T^{1\gamma^-}_{N}(3)+\sum_{i=2}^{n-1}T^{1\gamma^-}_{N^2}(i+1) \right ) \,\label{1g:TN3}\,.
\eea

In Table \ref{nn2bcfw}, we list the number of terms of N$^{\frac{N}{2}}$MHV amplitudes for the process $0\to e^- e^+  \gamma_1^+ \cdots \gamma_n^+ \gamma_1^- \cdots \gamma_n^-$ in the LLYZ shift, $[\gamma^-, \gamma^- \rangle$ shift, and the $[1, \gamma^- \rangle$ shift in the BCFW method. It is noticed that the number of terms in the full amplitudes in LLYZ shift increases rapidly when compared with $[\gamma_1^-,\gamma_2^-\rangle$ and $[1^-,\gamma^+\rangle$ shifts. When $ N_\gamma  \geq 8$, the number of terms in the $[\gamma^-, \gamma^-\rangle$ has an obvious advantage.

It should be mentioned that the number of terms is not the only factor which determine the speed of computation. For different shifts in the BCFW method, the number of independent amplitudes, number of calculation to eliminate the shifted brackets, are also crucial to evaluate the speed of computation. 

\begin{table}[htb]
\centering
\begin{tabular}{|c|c|c|c|c|c|}
\hline
%Amplitudes  &  & Form Gauge independent\\
\hline
$N_p$ & 6 & 7 & 8 &  9 & 4+n \\
\hline
$N_\gamma^+$ & 2  & 3 & 4 &  5 & n \\
\hline
\hline
LLYZ shift   & 2  & 3  & 4 & 5  & n \\
\hline
$[\gamma_1^-,\gamma_2^-\rangle$  & 4 & 6 & 8 & 10  & 2 n \\
\hline
$[1^-,\gamma^+\rangle$  & 3  & 6  & 10 & 15 & $\frac{n (n+1)}{2}$ \\
\hline
\end{tabular}
\caption{Number of independent NMHV terms needed to be computed before using boson changing symmetries for $0\to e^- e^+  \gamma_1^+ \cdots \gamma_n^+ \gamma_1^- \gamma_2^-$ with a few typical shifts in the BCFW method are provided.}
\label{Tnn2bcfw}
\end{table}

\begin{table}[htb]
\centering
\begin{tabular}{|c|c|c|c|c|c|}
\hline
%Amplitudes  &  & Form Gauge independent\\
\hline
$N_p$ & 8 & 9 & 10 & 11 & 5+n \\
\hline
$N_\gamma^+$ & 3  & 4 & 5 & 6 & n \\
\hline
\hline
LLYZ shift   & 9  & 16  & 25 & 36  & $n^2$ \\
\hline
$[\gamma_i^-,\gamma_j^-\rangle$  & 18  & 32 & 50 & 72  & 2 $n^2$ \\
\hline
$[1^-,\gamma^+\rangle$  & 20  & 50  & 105 & 196 &$\frac{n}{6}+\frac{5n^2}{12}+\frac{n^3}{3}+\frac{n^4}{12}$ \\
\hline
\end{tabular}
\caption{Number of independent terms for $N^2MHV$ needed to be computed before using boson changing symmetries for $0\to e^- e^+  \gamma_1^+ \cdots \gamma_n^+ \gamma_1^- \gamma_2^-  \gamma_3^-$ with a few typical shifts in the BCFW method are provided.}
\label{Tnn3bcfw}
\end{table}
In Table \ref{Tnn2bcfw} and Table \ref{Tnn3bcfw}, we list the number of independent terms which should be computed before using the boson exchanging symmetries. For the NMHV method, LLYZ shift only have half number of independent terms to be computed. Meanwhile, the number of calculation to evaluate the shifted brackets in the LLYZ is much smaller than $[\gamma_i^-,\gamma_j^-\rangle$ shift. While the shift $[1^-,\gamma^+\rangle$ has the least number of terms in the full amplitudes, but there are much more independent terms should be computed before invoking the boson exchanging symmetry. This shift has more shifted brackets to be computed. 

\section{Discussions and Conclusions}
In this work, we have used the Feynman diagram method in BG gauge to obtain the NMHV amplitudes of $0 \to e^- e^+ 4 \gamma$. Especially, the NMHV amplitudes of $0 \to e^- e^+ 4 \gamma$ given in Eq. (\ref{nmhv}) can have a form with explicit boson exchanging symmetries. By using the amplitude, we have performed a comprehensive and detailed study on all allowed shifts. The equivalence of all allowed shifts are checked by the numerical method. It is interesting to note that the shift $[2, 1 \rangle$ can also yield the correct amplitude, which is not expected. Meanwhile, it is found that there exist  two pair of shifts,  which can lead to the full results in a new way, i.e. the sum of $ [  1, 5 \rangle  $ and $[ 1, 6  \rangle$ shifts ( or the sum of  $[ 3 , 2 \rangle $  and $[ 4 , 2 \rangle $ shifts ) can finally lead to the whole results. Naively, it is expected that each of them should be equal to the whole results. We have compared this novel shift with other shifts in the BCFW method. 
With tests for more amplitudes given in Eq. (\ref{nmhv7}), Eq.(\ref{nmhv8}) and Eq.(\ref{nmhv8n2}), we have demonstrated and proven this new shift (LLYZ shift) can be applied to the processes $0 \to e^- e^+ n \gamma$. It should be mentioned that even with non-vanishing boundary terms, the on-shell method can be used, as demonstrated in \cite{Feng:2009ei}.

It needs more work to expose analytically the results from different shifts in the BCFW method are equivalent even if we know two of them are equal numerically. An insightful observation is that the amplitudes can be viewed as the volume of polytopes defined in twistor space \cite{Hodges:2009hk}, which can be hold for NMHV and had been generalized to amplitudes of more general QFT and gravity theories \cite{Arkani-Hamed:2010wgm}. To explore the geometrical meanings of NMHV of $0 \to e^- e^+ 4 \gamma$ in QED will be interesting but beyond the scope of current work, which can be studied in our future works. 

As shown in Table \ref{nn2bcfw}, the increase of the number of terms of N$^{\frac{N}{2}}$MHV amplitudes in the LLYZ shift grows much faster than that of shift $[\gamma^-,\gamma^-\rangle$ and that of shift $[1^-, \gamma^+\rangle$,respectively. It is worthy to optimize this shift so as to avoid the number of terms to grow too quick with the increase of number of photons. It might also be interesting to examine whether such a shift can work for the process $0 \to q \bar{q} \, n g$ in the QCD. Besides the tree level amplitudes, it might be also interesting to examine whether the LLYZ shift can be applied to loop level amplitudes of YM gauge theories \cite{Boels:2011tp,Boels:2011mn}. 

\section*{Acknowledgements}
This work is supported by the Natural Science Foundation of China under the grant No. 11475180 and No. 11875260. X. Zhao's work is supported by the Italian Ministry of Research (MUR) under grant PRIN 20172LNEEZ.

\bibliographystyle{JHEP}
\bibliography{bibHel}
\end{document}